\documentclass[aps,pra,print,twocolumn,groupedaddress]{revtex4-2}
\usepackage{graphicx}
\usepackage{amssymb}
\usepackage{amsfonts}
\usepackage{amsmath}
\usepackage{natbib}
\usepackage{hyperref}
\usepackage{color}
\usepackage{bm}
\usepackage{mathrsfs}
\usepackage{amsthm}

\begin{document}

\title{Dark solitons in nonlinear Su-Schrieffer-Heeger lattices}
\author{Rujiang Li}
\thanks{Corresponding author: {rujiangli@xidian.edu.cn}}
\author{Muhammad Imran}
\author{Wencai Wang}
\author{Yongtao Jia}
\author{Ying Liu}

\affiliation{National Key Laboratory of Radar Detection and Sensing, School of Electronic
Engineering, Xidian University, Xi'an 710071, China}

\begin{abstract}

The introduction of nonlinearities into lattices with topological band structures has led to the discovery of various types of solitons. The Su-Schrieffer-Heeger (SSH) lattice, as the most fundamental topological model, has been extended into the nonlinear regime. In particular, nonlinear edge states and bulk solitons exhibiting intensity humps against a zero background have been extensively studied in nonlinear SSH lattices. In this paper, we systematically investigate dark solitons in nonlinear SSH lattices. These dark solitons maintain a nonzero and constant background, featuring intensity dips either in the bulk of the lattice or at its edges, and residing spectrally in the semi-infinite gap or the middle finite gap. Regardless of the specific type of dark soliton, the intensity dip remains well-preserved and is not affected by the band structure of the original linear lattice. 
Although the dark solitons we have identified are generally dynamically unstable across a broad range of parameters, several types exhibit linear stability when the intracell coupling is much larger than the intercell coupling.
Our findings may provide valuable insights for the exploration of novel types of solitons in nonlinear topological lattices.

\end{abstract}

\maketitle

\section{Introduction}

Topological insulators are condensed matter materials that function as conventional insulators in the bulk 
while maintaining surface conductivity through topologically protected edge states \cite{TI1,TI2,
RMP82-3045,RMP83-1057}. The concept of topological insulators, along with its generalizations \cite{QF3-14,
QF2-22,QF3-21}, has led to the realization of various counterparts across multiple physical platforms, such as 
acoustic and mechanical systems \cite{NRP1-281,NRM7-974,NRP5-483,RMP96-021002,QF3-26,RPP86-106501}, 
bosonic condensates in ultracold gases \cite{RMP91-015005,nphys12-639}, and photonics \cite{Segev,
RMP91-015006,LSA9-130,QF1-10,PIER,APR,PR1093-1,APLED,NSR8-nwaa192,PRR2-023180}. 
The immunity of topological edge states to local deformations and disorder is crucial for 
their promising potential applications. In the past decade, there has been a growing interest in exploring 
the new features that emerge when nonlinear effects are introduced into topological lattices \cite{APR7-021306,
NP20-905}. On one hand, novel types of nonlinear states, including solitons, may be discovered in nonlinear 
topological lattices \cite{PRA90-023813,PRA94-021801,optica3-1228,PRL119-253904,PRB102-115411,OL45-6466,
nphys18-678,PRL128-093901,PRE104-054206,CP8-342,PRL117-143901,ACSPhoton7-735,ncommun11-1902,
PRX11-041057,PRA103-053507,ACSPhoton8-1077,PRB106-195423,nphys17-995,PRL111-243905,PRL118-023901,
PRA98-013827,LPR13-1900223,arxiv1904-10312,science368-856,CP5-275,PRL123- 053902,
PRResearch5-L012041,PRB93-15512,nelectron1-178,ncommun10-1102,nnano14-2,PRA97-043602,
nano10-3559,CSF186-115239,NJP22-103058,arxiv-prb,CP8-451,CSF207-118044}. 
On the other hand, nonlinearity can be utilized to manipulate topological edge states \cite{PRL121-163901,
PRB104-235420,LSA9-147,science372-72,LSA10-164,nanophotonics14-769,PRL127-184101,PRL133-116602,
arXiv:2411.07522,ncommun16-422,nphys20-1164,nature596-63,PRL128-154101,PRL128-113901,ncommun13-5997,
nphys19-420,science384-317,OL51-1649}.

The Su-Schrieffer-Heeger (SSH) lattice is the simplest yet most prominent type of topological lattice 
model \cite{PRL42-1698,book1}. By replacing the original constant couplings with intensity-dependent ones, 
self-induced topological edge states that decay to a non-zero plateau level have been 
discovered \cite{PRB93-15512,nelectron1-178}. 
Meanwhile, topologically enhanced harmonic generation can be achieved in these nonlinear SSH 
lattices \cite{ncommun10-1102,nnano14-2}. Another approach to constructing nonlinear SSH lattices involves 
using wavefunction-dependent onsite potentials. In these SSH lattices with onsite nonlinearity, 
nonlinear topological edge states, which originate from the linear counterpart, have been widely 
investigated \cite{PRB102-115411,OL45-6466,nphys18-678,PRE104-054206,LPR13-1900223,PRL121-163901,
PRB104-235420,LSA9-147,ncommun16-422,arxiv-prb}. 
Gap solitons, which spectrally reside within the topological bandgap, can be created in the bulk of 
the lattice \cite{PRB102-115411,OL45-6466,nphys18-678,PRL118-023901,LPR13-1900223,nanophotonics14-769,
arxiv-prb}. 
Notably, conventional solitons, which are topologically trivial states, also exist in the nonlinear SSH 
lattices \cite{PRB102-115411,nphys18-678,PRL118-023901,PRB104-235420,LSA9-147,arxiv-prb,
arxiv-prb}. Additionally, efforts to 
establish a general bulk-boundary correspondence in nonlinear SSH lattices are currently under 
investigation \cite{ncommun13-3379,FP18-33311,arxiv}.

In conventional soliton theory, in contrast to the bright solitons, dark solitons represent another prominent 
category, characterized by intensity dips within a continuous-wave background \cite{IEEEJQE29-250,
PR298-81}. By manipulating the dispersion of topological edge states, dark topological edge solitons have been 
proposed in two-dimensional and three-dimensional nonlinear topological lattices \cite{PRA97-043602,nano10-3559,
CSF186-115239,NJP22-103058}. However, as the most fundamental topological model,
the existence and properties of dark solitons in one-dimensional nonlinear SSH lattices have 
yet to be systematically explored. In particular, 
the previously studied nonlinear edge states and bulk solitons that reside in the topological bandgap become 
delocalized when they enter the linear bulk band \cite{PRB102-115411,PRL128-093901,CP8-342,nphys17-995,
PRB104-235420,arxiv-prb,CSF207-118044}. 
Since dark solitons are inherently delocalized states, it is important to investigate whether the intensity dips can always be effectively preserved.

In this paper, we systematically investigate dark solitons in nonlinear SSH lattices. In the anti-continuum (AC) limit, where intercell couplings vanish, a single unit cell supports two types of solutions with symmetric and antisymmetric amplitude distributions. Building on these solutions, we derive various types of dark solitons that feature a nonzero background and exhibit intensity dips either in the bulk of the lattice or at its edges. Notably, to ensure a constant soliton background, even at the lattice edges, we consider variants of the nonlinear SSH lattices with modified boundary conditions. 
For dark bulk solitons exhibiting intensity dips in the bulk of the lattice, regardless of whether the original linear SSH lattice is topologically nontrivial or trivial, those constructed from the symmetric solution in the AC limit consistently reside within the semi-infinite gap of the band structure, while those derived from the antisymmetric solution occupy both the semi-infinite gap and the middle finite gap. For dark edge solitons with intensity dips at the lattice edges, if the original linear SSH lattice is topologically nontrivial, solitons constructed from either the symmetric or antisymmetric solution in the AC limit consistently reside within the semi-infinite gap. In contrast, if the original SSH lattice is topologically trivial, dark edge solitons constructed from the antisymmetric solution may exist in both the semi-infinite gap and the middle finite gap, while those constructed from the symmetric solution exist only within the semi-infinite gap. Regardless of the specific type of dark soliton, the intensity dip remains well-preserved whether the frequency is situated in the middle finite gap, the semi-infinite gap, or the band corresponding to the linear bulk states. Although the dark solitons we have identified are generally unstable across a broad range of parameters, several types exhibit linear stability when the intercell coupling is much smaller than the intracell coupling. Our work may provide valuable insights for exploring novel types of solitons in nonlinear topological lattices.

This paper is organized as follows. In Section II, we introduce the nonlinear SSH lattice model and provide a brief review of the band structure of the original linear lattice. In Section III, using the approach of the AC limit, we investigate the dark solitons residing in the bulk of the originally topologically nontrivial lattice, examining both their frequency spectra and stability properties. Following a similar approach, in Section IV, we study the dark solitons in the bulk of the originally topologically trivial lattice. In Sections V and VI, we explore the dark solitons located at the edges of the originally 
topologically nontrivial and trivial lattices, respectively, and also discuss the frequency spectra and stability properties 
of the dark edge solitons. In Section VII, we discuss our results and compare them with previous studies on dark solitons.
Finally, Section VIII summarizes the main findings of this paper.

\section{Model of nonlinear Su-Schrieffer-Heeger (SSH) lattices}

\begin{figure}[tbp]
\includegraphics[width=8.6cm]{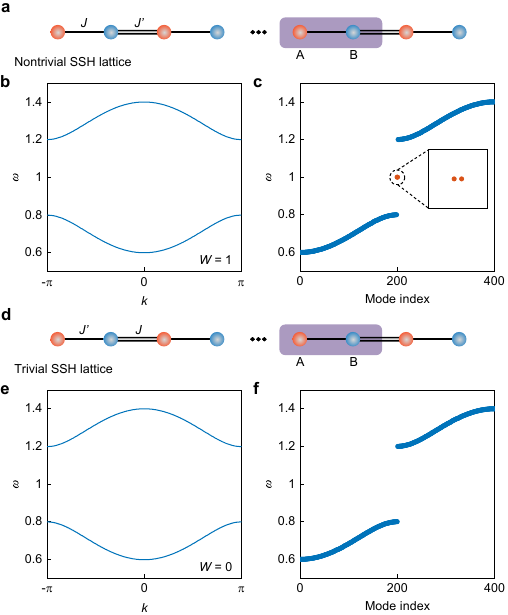}
\caption{\textbf{Su-Schrieffer-Heeger (SSH) lattices and their corresponding band structures and frequency spectra.} 
\textbf{a} Geometry of the topologically nontrivial SSH lattice. 
\textbf{b} Band structure of the nontrivial SSH lattice with periodic boundaries. 
\textbf{c} Frequency spectrum of the nontrivial SSH lattice with open boundaries. 
\textbf{d}-\textbf{f} Geometry, band structure, and frequency spectrum of the topologically trivial SSH lattice, respectively. 
In (c) and (f), the topological edge states are indicated by red dots, while the bulk states are represented by blue dots. 
Throughout all panels, we maintain the condition $J < J'$.}
\label{fig_linear}
\end{figure}

For completeness, we begin with a brief review of the results for linear SSH lattices. The Hamiltonian for the linear SSH lattice model is given by
\begin{eqnarray}
\hat{H} &=&E_{0}\sum_{n} \left( c_{n}^{\text{A}\dag }c_{n}^{\text{A}} + c_{n}^{\text{B}\dag }c_{n}^{\text{B}} \right)
-J\sum_{n} \left( c_{n}^{\text{B}\dag }c_{n}^{\text{A}}+\text{H.c.} \right)  \notag \\
&&-J^{\prime}\sum_{n}  \left(c_{n+1}^{\text{A}\dag }c_{n}^{\text{B}} +\text{H.c.}\right),    \label{H_SSH}
\end{eqnarray}
where $E_{0}$ is the onsite potential, $c_{n}^{\sigma\dag }$ and $c_{n}^{\sigma}$ are the creation and
annihilation operators for the particles at the sublattice site $\text{A}$ ($\sigma = \text{A}$) 
or $\text{B}$ ($%
\sigma = \text{B}$) of the lattice cell $n$, the summation enumerated with the index $n$
is performed over all the cells in the lattice, and $\text{H.c.}$ denotes the Hermitian
conjugate of the term to the left \cite{PRL42-1698,book1}. The Hamiltonian given by Eq. (\ref{H_SSH})
describes particle hopping along a one-dimensional chain. As illustrated in Fig. \ref{fig_linear}a, 
the hopping amplitudes are staggered, with intracell hopping $J$ (single lines) differing from intercell hopping $J^{\prime}$ (double lines). 
The SSH model can be realized using a lattice composed of coupled resonators, and the dynamics of the lattice obey the following equation:
\begin{equation}
\mathrm{i}\frac{d}{dt}%
\begin{bmatrix}
\psi _{n}^{\mathrm{A}} \\ 
\psi _{n}^{\mathrm{B}}%
\end{bmatrix}%
=\omega _{0}%
\begin{bmatrix}
\psi _{n}^{\mathrm{A}} \\ 
\psi _{n}^{\mathrm{B}}%
\end{bmatrix}%
+J%
\begin{bmatrix}
\psi _{n}^{\mathrm{B}} \\ 
\psi _{n}^{\mathrm{A}}%
\end{bmatrix}%
+J^{\prime}%
\begin{bmatrix}
\psi _{n-1}^{\mathrm{B}} \\ 
\psi _{n+1}^{\mathrm{A}}%
\end{bmatrix},
\label{eq_linear}
\end{equation}
where $\omega_{0}$ is the resonant frequency of a single resonator, and $J$ and $J^{\prime}$ are the intracell 
and intercell coupling coefficients, respectively \cite{NSR8-nwaa192,CP8-342,PRB93-15512,nelectron1-178,
arxiv-prb,PRL133-116602,PRA103-023503,iscience}. When the SSH lattice has periodic boundaries, 
the solution of Eq. (\ref{eq_linear}) can be expressed in the form of Bloch states:
$\psi_{n}^{\mathrm{A},\mathrm{B}} = \phi_{\mathrm{A},\mathrm{B}} \exp\left(\mathrm{i} k n -
\mathrm{i} \omega t\right)$,
where $\phi_{\mathrm{A},\mathrm{B}}$ are the Bloch states, $k$ is the wavenumber, and
$\omega$ is the frequency. The band structure which describes the dependence between $\omega$ and $k$,
is expressed as $\omega= \omega_{0} \pm \sqrt{J^2+J^{\prime2}+2J J^{\prime}\cos k}$.
As illustrated in Fig.~\ref{fig_linear}b, when $J < J^{\prime}$, the band structure of the SSH lattice exhibits a finite bandgap in the middle. 
The edges of the bulk states are given by $\omega_{0} - J - J^{\prime}$, $\omega_{0} + J - J^{\prime}$, $\omega_{0} - J + J^{\prime}$, 
and $\omega_{0} + J + J^{\prime}$, respectively, from bottom to top.
Meanwhile, the lattice is topologically nontrivial, characterized by a winding number given by
$W=\frac{\mathrm{i}}{\pi}\int_{-\pi}^{\pi}\left\langle \phi\right\vert
\frac{\partial}{\partial k}\left\vert \phi\right\rangle dk = 1$, where 
$\left\vert \phi \right\rangle= \left[\phi_{\mathrm{A}}, \phi_{\mathrm{B}} \right]^{T}$
with $T$ denoting the transpose of the vector \cite{book1,PRB84-195452}.
Due to the bulk-boundary correspondence, when the lattice has open boundaries, it supports two topological edge states (shown as red dots in Fig. \ref{fig_linear}c) located at the two ends of the lattice, in addition to the bulk states (denoted by blue dots). 
The parameters for the linear SSH lattice referenced in Figs.~\ref{fig_linear}a-c are as follows: 
$\omega_{0}=1$, $J=0.1$, and $J^{\prime}=0.3$. Additionally, the lattice shown in Fig.~\ref{fig_linear}c contains $N=200$ unit cells.

We maintain the condition $J < J^{\prime}$ and fix the values of all parameters while exchanging the intracell and intercell couplings. The new lattice is schematically illustrated in Fig. \ref{fig_linear}d, where $J^{\prime}$ and $J$ correspond to the intracell and intercell coupling coefficients, respectively. Under periodic boundary conditions, Fig. \ref{fig_linear}e shows that this lattice exhibits an identical band structure to that presented in 
Fig. \ref{fig_linear}b. Although it also features a finite bandgap, the lattice is topologically trivial, with a winding number of $W = 0$. Therefore, 
the lattice with open boundaries does not support topological edge states. In this case, only bulk states are present, as indicated by the blue dots in 
Fig. \ref{fig_linear}f.

We then introduce onsite nonlinear terms to Eq. (\ref{eq_linear}), resulting in the following new equations:
\begin{eqnarray}
\mathrm{i}\frac{d\psi _{n}^{\mathrm{A}}}{dt} &=& \omega_{0} \psi_{n}^{\mathrm{A}} + J \psi_{n}^{\mathrm{B}} + J^{\prime} \psi_{n-1}^{\mathrm{B}} + \gamma \left| \psi_{n}^{\mathrm{A}} \right|^{2} \psi_{n}^{\mathrm{A}}, 
\label{eq1}\\
\mathrm{i}\frac{d\psi _{n}^{\mathrm{B}}}{dt} &=& \omega_{0} \psi_{n}^{\mathrm{B}} + J \psi_{n}^{\mathrm{A}} + J^{\prime} \psi_{n+1}^{\mathrm{A}} + \gamma \left| \psi_{n}^{\mathrm{B}} \right|^{2} \psi_{n}^{\mathrm{B}},
\label{eq2}
\end{eqnarray}
where $\gamma$ is the Kerr nonlinear coefficient. In the following sections, we will seek dark soliton solutions of Eqs. (\ref{eq1})-(\ref{eq2}), 
which correspond to dark solitons in the originally topologically nontrivial lattice. For dark solitons in the originally topologically trivial lattice, 
we simply need to exchange $J$ and $J^{\prime}$ in Eqs. (\ref{eq1})-(\ref{eq2}). We will focus on dark solitons with intensity dips in the bulk of the lattice, 
as well as those with intensity dips at the edges, i.e., dark bulk solitons and dark edge solitons. Since $\psi_n^{\mathrm{A,B}}$ can 
always be normalized by the replacement $\psi_n^{\mathrm{A,B}} / \sqrt{|\gamma|}$, we will set $\gamma = 1$ in the following study without 
loss of generality. When $\gamma < 0$, a phase difference of $\pi$ between the field amplitudes at sites $\mathrm{A}$ and $\mathrm{B}$ can be 
introduced. Note that the constant onsite terms can be eliminated by the transformation $\psi_n^{\mathrm{A,B}} \to \psi_n^{\mathrm{A,B}} \exp \left(-\mathrm{i} \omega_0 t \right)$.

\section{Dark bulk solitons in the originally topologically nontrivial lattice}

For the dark bulk solitons in a given nonlinear SSH lattice, the intensity dips can occur at either two sites within a single unit cell or at neighboring sites across two different unit cells. However, since the intracell and intercell couplings can be interchanged to transform an originally topologically nontrivial lattice into an originally topologically trivial one, it suffices to study the dark bulk solitons with intensity dips located at the two sites of a single unit cell in both originally topologically nontrivial and topologically trivial lattices.

\subsection{Solutions in the anti-continuum (AC) limit}

In this study, we focus on dark solitons that arise as continuations of solutions in the AC limit, which represents the extreme case of a lattice with no intersite hopping \cite{AC_limit_1,AC_limit_2}.
We set $J' = 0$, indicating that the intercell coupling in the originally topologically nontrivial lattice is eliminated, resulting in decoupled unit cells within the nonlinear SSH lattice. 
With $J^{\prime}=0$ and $\psi_n^{\mathrm{A},\mathrm{B}} = \phi_n^{\mathrm{A},\mathrm{B}} \exp(-\mathrm{i} \omega t)$, where
$\omega$ is the frequency, Eqs. (\ref{eq1})-(\ref{eq2}) reduce to
\begin{eqnarray}
\omega \phi_n^{\mathrm{A}} &=& \omega_0 \phi_n^{\mathrm{A}} + J \phi_n^{\mathrm{B}} + \gamma \left\vert \phi_n^{\mathrm{A}} \right\vert^2 \phi_n^{\mathrm{A}}, 
\label{AC_eq1} \\
\omega \phi_n^{\mathrm{B}} &=& \omega_0 \phi_n^{\mathrm{B}} + J \phi_n^{\mathrm{A}} + \gamma \left\vert \phi_n^{\mathrm{B}} \right\vert^2 \phi_n^{\mathrm{B}}, \label{AC_eq2}
\end{eqnarray}
which describe the wave dynamics in a single unit cell composed of two coupled resonators. 
Typically, in addition to the trivial zero solution, there are three branches of nonzero solutions: one branch corresponds to symmetric solutions, 
another to antisymmetric solutions, and the last to doubly degenerate asymmetric solutions \cite{PD,OL51-1649}. For simplicity, we will only 
consider the symmetric solution given by
$\phi_n^{\mathrm{A}} = \phi_n^{\mathrm{B}} = \pm \sqrt{\left(\omega - \omega_0 - J\right)/\gamma}$
and the antisymmetric solution given by
$\phi_n^{\mathrm{A}} = -\phi_n^{\mathrm{B}} = \pm \sqrt{\left(\omega - \omega_0 + J\right)/\gamma}$.
From the two solutions, the existence of real-valued symmetric and antisymmetric solutions requires $\omega > \omega_0 + J$ and 
$\omega > \omega_0 - J$, respectively. We can expect that when the frequency $\omega$ lies within the upper semi-infinite gap shown in 
Fig. \ref{fig_linear}b, both dark solitons that continue from the symmetric solution and those that continue from the antisymmetric solution 
should exist. In contrast, when the frequency lies within the middle finite gap, at least the dark bulk solitons that continue from the antisymmetric 
solution may exist.

Based on the solutions for a single unit cell, we construct the solutions for dark bulk solitons in the AC limit of $J^{\prime} = 0$. 
Utilizing the symmetric solution, we can express the dark bulk solitons as 
\begin{equation}
\phi _{n}^{\mathrm{A}} = \phi _{n}^{\mathrm{B}} = \sqrt{\frac{\omega - \omega_{0} - J}{\gamma}} 
\left\vert \mathrm{sgn} \left(n\right) \right\vert,
\label{solu_nontrivial_sym_1}
\end{equation}
or 
\begin{equation}
\phi _{n}^{\mathrm{A}} = \phi _{n}^{\mathrm{B}} = \sqrt{\frac{\omega - \omega_{0} -J}{\gamma}} 
\mathrm{sgn} \left(n\right),
\label{solu_nontrivial_sym_2}
\end{equation}
where the sign function $\mathrm{sgn}(n)$ is defined as follows: it equals $1$ for $n \geq 1$, $-1$ for $n \leq -1$, and $0$ when $n = 0$.
Both types of solutions possess a nonzero intensity background and exhibit intensity dips in the bulk of the lattice. However, the latter solution 
features a phase jump when crossing the intensity dip, while the former does not. Conventionally, 
the solution given by Eq. (\ref{solu_nontrivial_sym_1}) is referred to as the bubble soliton \cite{bubble_soliton,PD34-240}. 
In this paper, we collectively refer to both solutions provided by 
Eqs. (\ref{solu_nontrivial_sym_1})-(\ref{solu_nontrivial_sym_2}) as dark bulk solitons.
Similarly, using the antisymmetric solution, we can construct the dark bulk solitons as 
\begin{equation}
\phi _{n}^{\mathrm{A}} = -\phi _{n}^{\mathrm{B}} = \sqrt{\frac{\omega - \omega_{0} + J}{\gamma}} 
\left\vert \mathrm{sgn} \left(n\right) \right\vert,
\label{solu_nontrivial_anti_1}
\end{equation}
or 
\begin{equation}
\phi _{n}^{\mathrm{A}} = -\phi _{n}^{\mathrm{B}} = \sqrt{\frac{\omega - \omega_{0} + J}{\gamma}} 
\mathrm{sgn} \left(n\right).
\label{solu_nontrivial_anti_2}
\end{equation}
These two types of solutions also have a nonzero intensity background and exhibit intensity dips in the bulk of the lattice. 
It is noteworthy that the intensity background of all four solutions for dark bulk solitons, as shown in 
Eqs. (\ref{solu_nontrivial_sym_1})-(\ref{solu_nontrivial_anti_2}), extends into both positive and negative infinities. 
In other words, these dark bulk solitons are inherently delocalized states.

\begin{figure}[tbp]
\includegraphics[width=8.6cm]{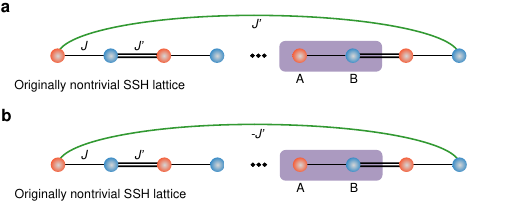}
\caption{\textbf{The originally topologically nontrivial SSH lattices with modified boundaries employed to seek dark bulk solitons.}  
\textbf{a} Nonlinear SSH lattice where the leftmost site is coupled to the rightmost one with the coupling coefficient $J^{\prime}$. 
This lattice configuration accommodates the dark bulk solitons that are continuations of the solutions given by Eqs. (\ref{solu_nontrivial_sym_1}) and (\ref{solu_nontrivial_anti_1}).  
\textbf{b} Nonlinear SSH lattice where the leftmost site is coupled to the rightmost one with the coupling coefficient $-J^{\prime}$. 
This lattice configuration ensures the phase jump required by the dark bulk solitons that continue from the solutions given 
by Eqs. (\ref{solu_nontrivial_sym_2}) and (\ref{solu_nontrivial_anti_2}).}
\label{fig_structure_bulk}
\end{figure}

\subsection{Existence of dark bulk solitons}

\begin{figure}[tbp]
\includegraphics[width=8.6cm]{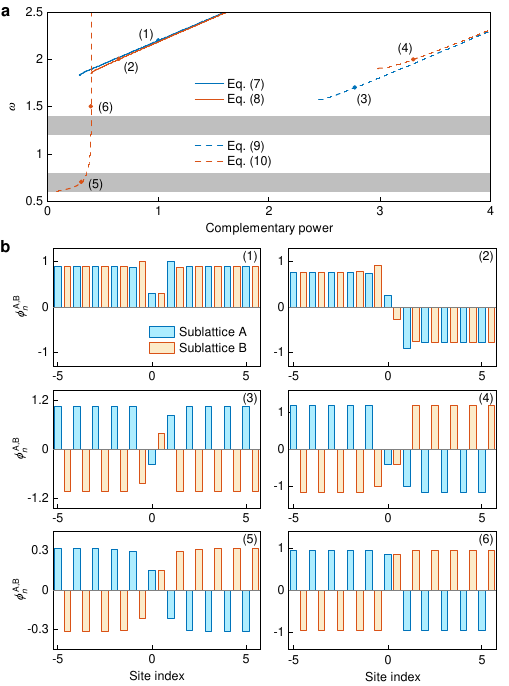}
\caption{\textbf{The existence of dark bulk solitons in the originally topologically nontrivial lattice.}
\textbf{a} Existence curves of the dark bulk solitons, showing the dependence between frequency and complementary power. 
The dark bulk solitons, which are continuations of distinct solutions in the AC limit, are labeled in the legend for better understanding. 
The gray regions denote the frequency bands associated with the bulk states of the originally 
linear SSH lattice.
\textbf{b} Amplitude distributions of the dark bulk solitons, which are marked on the existence curves in (a). 
The blue and red bars represent the amplitudes at sublattice A and sublattice B, respectively. 
For clarity, we only show the soliton amplitudes at several sites in the middle of the lattice.}
\label{fig_nontrivial_bulk_mode}
\end{figure}

Based on the four solutions in the AC limit of $J' = 0$, as shown in Eqs. (\ref{solu_nontrivial_sym_1})-(\ref{solu_nontrivial_anti_2}), 
we can utilize Newton's method to find dark bulk solitons in the nonlinear SSH lattice by gradually increasing $J'$ from zero to the preset value of $J^{\prime} = 0.3$. In numerical calculations, 
if open boundary conditions are applied directly, the soliton solutions become perturbed by the boundaries of the lattice, leading to 
the intensity background deviating from a constant. To ensure that the solutions for dark bulk solitons remain consistent with the 
ideal dark solitons that maintain a constant background, even at the lattice edges, we must implement the nonlinear SSH lattice with appropriately modified boundary conditions.
For the solutions given by Eqs. (\ref{solu_nontrivial_sym_1}) and (\ref{solu_nontrivial_anti_1}), $\phi_{n}^{\mathrm{A},\mathrm{B}}$ is inversion symmetric with respect to the middle unit cell at $n=0$. In this case, we can connect the leftmost lattice site to the rightmost one using the coupling coefficient $J^{\prime}$, as schematically illustrated in Fig. \ref{fig_structure_bulk}a. In contrast, for the solutions given by Eqs. (\ref{solu_nontrivial_sym_2}) and (\ref{solu_nontrivial_anti_2}), $\phi_{n}^{\mathrm{A},\mathrm{B}}$ is inversion antisymmetric with respect to the middle unit cell. In this scenario, the leftmost lattice site can be connected to the rightmost one with the coupling coefficient $-J^{\prime}$, as shown in Fig. \ref{fig_structure_bulk}b. The implementation of this negative coupling ensures a phase jump of the dark bulk solitons when crossing the intensity dip.

Figure \ref{fig_nontrivial_bulk_mode}a displays the families of different types of dark bulk solitons in the originally topologically nontrivial SSH lattice with modified boundaries. 
The dark bulk solitons, which are continuations of distinct solutions in the AC limit, are labeled in the legend for better understanding.
For comparison, the frequency bands associated with the linear bulk states shown in Figs. \ref{fig_linear}b-c are represented by the gray regions. To characterize the dark bulk solitons, we adopt the conventional definition \cite{IEEEJQE29-250,PR298-81}, in which the complementary power is defined as
\begin{equation}
P_{r} = 2N \vert \phi_{\mathrm{b}} \vert^{2} - \sum_{n} \left( \left\vert \phi_{n}^{\mathrm{A}} \right\vert^{2} 
+ \left\vert \phi_{n}^{\mathrm{B}} \right\vert^{2} \right), \label{Pr}
\end{equation}
where $\phi_{\mathrm{b}}$ is the amplitude of the background. 
Note that all the soliton branches terminate at low frequencies due
to the inability of the Newton’s method to capture poorly localized solutions.
Figure \ref{fig_nontrivial_bulk_mode}b displays the amplitude distributions of the dark bulk solitons, with their corresponding frequencies indicated on the existence curves shown in Fig. \ref{fig_nontrivial_bulk_mode}a. 
The site index is defined as $x = n$ for the sublattice site $\mathrm{A}$ and $x = n + 0.5$ for the sublattice $\mathrm{B}$. 
The amplitudes at sublattice A and sublattice B are represented by the blue and red bars, respectively. 
It is noteworthy that although we only show the soliton amplitudes at several sites in the middle of the lattice, all the dark bulk solitons illustrated in Fig. \ref{fig_nontrivial_bulk_mode}b exhibit constant backgrounds. In other words, the amplitude distributions of the dark bulk solitons are unaffected by the lattice edges. This constant background achieved through modified boundary conditions ensures the accurate evaluation 
of the complementary power.

As shown in Fig. \ref{fig_nontrivial_bulk_mode}a, the continuations of Eqs. (\ref{solu_nontrivial_sym_1})--(\ref{solu_nontrivial_anti_1}) yield distinct branches of dark bulk solitons, with all three types residing spectrally within the semi-infinite gap. In addition to these types, there are two more isolated branches, both originating from Eq. (\ref{solu_nontrivial_anti_2}). One branch lies within the semi-infinite gap, while the other extends from the lower edge of the bottom band of the original linear SSH lattice to positive infinity.
As the frequency $\omega$ increases, the complementary power of the dark bulk solitons in the branch corresponding to states (5) and (6) exhibits saturation, while the soliton intensity only slightly decreases at the two central sites of the lattice. It is important to note that although this branch 
is continuous, only the solitons in the segment with $\omega_{0} - J < \omega < \omega_{0} + J$ are continuations of the solutions given by Eq. (\ref{solu_nontrivial_anti_2}).
Furthermore, regardless of whether the frequency lies within the middle finite gap, the semi-infinite gap, or the band of linear bulk states, the intensity dips of all the dark bulk solitons shown in Fig. \ref{fig_nontrivial_bulk_mode} remain well preserved. In contrast, bright bulk solitons become delocalized when the frequency enters the band of linear bulk states, which results in the destruction of the intensity humps \cite{PRB102-115411,PRL128-093901,CP8-342,nphys17-995,PRB104-235420,arxiv-prb,CSF207-118044}.

\subsection{Stability analysis of dark bulk solitons}

\begin{figure}[tbp]
\includegraphics[width=8.6cm]{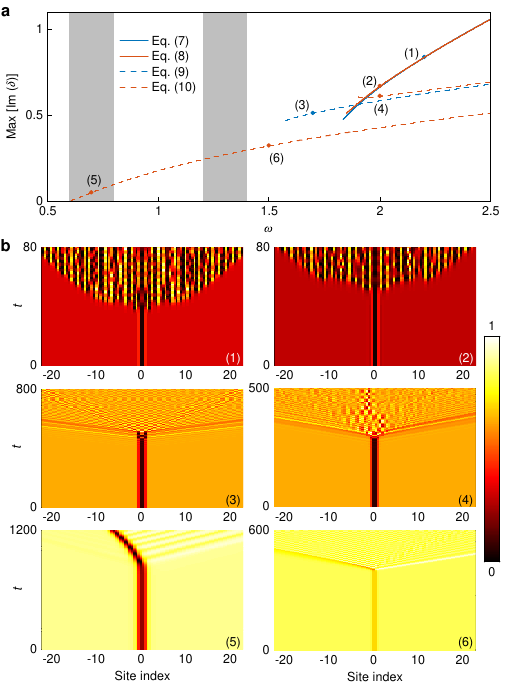}
\caption{\textbf{The stability analysis of dark bulk solitons in the originally topologically nontrivial lattice.}
\textbf{a} Maximum growth rates for the perturbed dark bulk solitons at various frequencies. The dark bulk solitons, 
which are continuations of distinct solutions in the AC limit, are labeled in the legend. 
The gray regions represent the frequency bands associated with the bulk states of the original linear SSH lattice.
\textbf{b} Temporal evolution of dark bulk solitons selected at specific frequencies indicated on the curves in (a).
For comparison across different cases, the intensities are normalized to their respective maximum values, 
with black representing zero and white representing one. }
\label{fig_nontrivial_bulk_stability}
\end{figure}

We then address the stability analysis of dark bulk solitons illustrated in Fig. \ref{fig_nontrivial_bulk_mode}. Following the standard procedure for linear stability analysis, the perturbed solutions of Eqs. (\ref{eq1})-(\ref{eq2}) are given by
\begin{equation}
\psi _{n}^{\sigma}(t) = e^{-\mathrm{i}\omega t}\left( \phi _{n}^{\sigma} + \varepsilon_{n}^{\sigma} e^{-\mathrm{i}\delta t} 
+ \mu _{n}^{\sigma \ast} e^{\mathrm{i}\delta^{\ast} t}\right),
\label{per_solu}
\end{equation}
with $\sigma =\mathrm{A}, \mathrm{B}$. Here,
$\varepsilon_{n}^{\mathrm{A,B}}$ and $\mu_{n}^{\mathrm{A,B}}$ are the infinitesimal perturbations. The dark bulk solitons are linearly stable if the eigenvalue $\delta$ is real; conversely, they are linearly unstable if the imaginary part of $\delta$, known as the growth rate, is positive. 
We substitute the perturbed solutions given by Eq. (\ref{per_solu})
into Eqs. (\ref{eq1})-(\ref{eq2}) and carry out the linearization by neglecting the higher-order terms of the perturbations. 
The resulting linearized equations for $\varepsilon_{n}^{\mathrm{A,B}}$ and $\mu_{n}^{\mathrm{A,B}}$ are given by
\begin{eqnarray}
\Delta \omega_{\mathrm{A}} \varepsilon_{n}^{\mathrm{A}} + J \varepsilon_{n}^{\mathrm{B}} + J^{\prime} \varepsilon_{n-1}^{\mathrm{B}} + \gamma \phi_{n}^{\mathrm{A}2} \mu_{n}^{\mathrm{A}} &=& \delta \varepsilon_{n}^{\mathrm{A}}, \\
-\Delta \omega_{\mathrm{A}} \mu_{n}^{\mathrm{A}} - J \mu_{n}^{\mathrm{B}} - J^{\prime} \mu_{n-1}^{\mathrm{B}} - \gamma \phi_{n}^{\mathrm{A} \ast 2} \varepsilon_{n}^{\mathrm{A}} &=& \delta \mu_{n}^{\mathrm{A}}, \\
\Delta \omega_{\mathrm{B}} \varepsilon_{n}^{\mathrm{B}} + J \varepsilon_{n}^{\mathrm{A}} + J^{\prime} \varepsilon_{n+1}^{\mathrm{A}} + \gamma \phi_{n}^{\mathrm{B}2} \mu_{n}^{\mathrm{B}} &=& \delta \varepsilon_{n}^{\mathrm{B}}, \\
-\Delta \omega_{\mathrm{B}} \mu_{n}^{\mathrm{B}} - J \mu_{n}^{\mathrm{A}} - J^{\prime} \mu_{n+1}^{\mathrm{A}} - \gamma \phi_{n}^{\mathrm{B} \ast 2} \varepsilon_{n}^{\mathrm{B}} &=& \delta \mu_{n}^{\mathrm{B}}.
\end{eqnarray}
Here, $\Delta \omega_{\mathrm{A,B}} = \omega_{0} - \omega + 2\gamma \vert \phi_{n}^{\mathrm{A,B}} \vert ^{2}$ 
is defined. Note that the boundary conditions shown in Fig. \ref{fig_structure_bulk} are also applied here.

Figure \ref{fig_nontrivial_bulk_stability}a shows the maximum growth rates, $\mathrm{Max} \left[ \mathrm{Im} \left( \delta \right) \right]$, 
for the perturbed dark bulk solitons at various frequencies. For comparison, the frequency bands corresponding to the original linear SSH lattice are indicated by the gray regions.
Notably, since $\mathrm{Max} \left[ \mathrm{Im} \left( \delta \right) \right] > 0$ for all dark bulk solitons across all frequencies, this indicates that all the dark bulk solitons shown in Fig. \ref{fig_nontrivial_bulk_mode} are linearly unstable. 
As the frequency $\omega$ approaches the lower edge of the bottom band of the original linear SSH lattice, the maximum growth rate of the dark bulk solitons in the branch corresponding to states (5) and (6) approaches zero.
This suggests that this type of dark bulk solitons become weakly unstable in this limit.

To corroborate the results from the linear stability analysis,
we perform direct simulations of the temporal evolution of the dark bulk solitons selected at specific frequencies 
indicated on the curves in Fig. \ref{fig_nontrivial_bulk_stability}a, using the Runge-Kutta algorithm.
The boundary conditions are consistent with those illustrated in Fig. \ref{fig_structure_bulk}.
No extra random perturbations are added to the initial excitations of states (1)-(6). 
As shown in Fig. \ref{fig_nontrivial_bulk_stability}b, the intensities of all types of dark bulk solitons break up after different durations of time. Here, the intensities are defined as $\left\vert \psi_{n}^{\mathrm{A,B}} \right\vert ^{2} / 
\mathrm{Max} \left( \left\vert \psi_{n}^{\mathrm{A,B}} \right\vert ^{2} \right)$, meaning that the intensities in different cases are normalized with 
respect to their respective maxima. The break up of intensity during temporal evolution implies that the dark bulk solitons 
are dynamically unstable, which is consistent with the results obtained from the linear stability analysis.

We would like to note that all the results mentioned above, including the frequency spectra and stability properties of dark bulk solitons, 
remain unchanged when we consider a larger lattice with $N = 400$ unit cells. This suggests that the current parameter 
value of $N = 200$ is sufficiently large and suitable for our study of dark solitons in this work.

Besides, we have also calculated different types of dark bulk solitons for various values of $J$ and $J^{\prime}$, 
ensuring that $J + J^{\prime} = 0.4$ and $J/J^{\prime} < 1$. The latter condition indicates that the originally linear SSH lattice is topologically nontrivial.
The dark bulk solitons that continue from Eqs. (\ref{solu_nontrivial_sym_1})-(\ref{solu_nontrivial_sym_2}) consistently reside in 
the semi-infinite gap and are always linearly unstable. In contrast, the dark bulk solitons that originate from Eq. (\ref{solu_nontrivial_anti_1}) 
can enter the band of linear bulk states 
when the ratio $J/J^{\prime}$ approaches one. This assertion will be further supported 
in our study of dark bulk solitons in the originally topologically trivial lattice.
Moreover, these dark bulk solitons remain consistently linearly unstable. For the dark bulk solitons that continue from Eqs. (\ref{solu_nontrivial_anti_2}), 
the branch corresponding to state (4) always resides in the semi-infinite gap, and the dark bulk solitons in this branch are linearly unstable. 
The branch corresponding to states (5) and (6) occupies the same position as that shown in Fig. \ref{fig_nontrivial_bulk_mode}a, and the dark bulk solitons in this branch are also linearly unstable.

\section{Dark bulk solitons in the originally topologically trivial lattice}

\subsection{Solutions in the AC limit}

For the originally topologically trivial lattice, we examine the solutions in the AC limit with $J = 0$, 
indicating that the intercell coupling is eliminated.
By exchanging $J$ and $J'$ in Eqs. (\ref{eq1})-(\ref{eq2}), we obtain the following equations:
\begin{eqnarray}
\mathrm{i}\frac{d\psi _{n}^{\mathrm{A}}}{dt} &=& \omega_{0} \psi_{n}^{\mathrm{A}} + J' \psi_{n}^{\mathrm{B}} + J \psi_{n-1}^{\mathrm{B}} + \gamma \left\vert \psi_{n}^{\mathrm{A}}\right\vert^2 \psi_{n}^{\mathrm{A}}, \label{eq_trivial_1} \\
\mathrm{i}\frac{d\psi _{n}^{\mathrm{B}}}{dt} &=& \omega_{0} \psi_{n}^{\mathrm{B}} + J' \psi_{n}^{\mathrm{A}} + J \psi_{n+1}^{\mathrm{A}} + \gamma \left\vert \psi_{n}^{\mathrm{B}}\right\vert^2 \psi_{n}^{\mathrm{B}}. \label{eq_trivial_2}
\end{eqnarray}
With $J=0$ and $\psi_{n}^{\mathrm{A},\mathrm{B}} = \phi_{n}^{\mathrm{A},\mathrm{B}} \exp(-\mathrm{i} \omega t)$, Eqs. (\ref{eq_trivial_1})-(\ref{eq_trivial_2}) reduce to:
\begin{eqnarray}
\omega \phi_{n}^{\mathrm{A}} &=& \omega_{0} \phi_{n}^{\mathrm{A}} + J' \phi_{n}^{\mathrm{B}} + \gamma \left\vert \phi_{n}^{\mathrm{A}}\right\vert^2 \phi_{n}^{\mathrm{A}}, \label{eqs_trivial_1}\\
\omega \phi_{n}^{\mathrm{B}} &=& \omega_{0} \phi_{n}^{\mathrm{B}} + J' \phi_{n}^{\mathrm{A}} + \gamma \left\vert \phi_{n}^{\mathrm{B}}\right\vert^2 \phi_{n}^{\mathrm{B}}. \label{eqs_trivial_2}
\end{eqnarray}
The symmetric solution of Eqs. (\ref{eqs_trivial_1})-(\ref{eqs_trivial_2}) is given by
$\phi_{n}^{\mathrm{A}} = \phi_{n}^{\mathrm{B}} = \pm \sqrt{\left(\omega - \omega_{0} - J^{\prime}\right)
/\gamma}$,
and the antisymmetric solution is given by 
$\phi_{n}^{\mathrm{A}} = -\phi_{n}^{\mathrm{B}} = \pm \sqrt{\left(\omega - \omega_{0} + J^{\prime}\right)
/\gamma}$.
From the two solutions, when the frequency $\omega$ lies within the upper semi-infinite gap shown in Fig. \ref{fig_linear}e, both the dark bulk solitons that continue from the symmetric solution and those that continue from the antisymmetric solution are expected to exist. When the frequency lies within the middle finite gap, at least the dark bulk 
solitons that continue from the antisymmetric solution are anticipated to exist.

Based on the symmetric solution for a single unit cell, the solutions for the whole lattice
can be constructed as%
\begin{equation}
\phi _{n}^{\mathrm{A}} = \phi _{n}^{\mathrm{B}} = \sqrt{\frac{\omega - \omega_{0} - J^{\prime}}{\gamma}} 
\left\vert \mathrm{sgn} \left(n\right) \right\vert,
\label{solu_trivial_sym_1}
\end{equation}
or 
\begin{equation}
\phi _{n}^{\mathrm{A}} = \phi _{n}^{\mathrm{B}} = \sqrt{\frac{\omega - \omega_{0} -J^{\prime}}{\gamma}} 
\mathrm{sgn} \left(n\right),
\label{solu_trivial_sym_2}
\end{equation}
Alternatively, based on the antisymmetric solution for a single unit cell, the solutions for
the whole lattice can be constructed as%
\begin{equation}
\phi _{n}^{\mathrm{A}} = \phi _{n}^{\mathrm{B}} = \sqrt{\frac{\omega - \omega_{0} + J^{\prime}}{\gamma}} 
\left\vert \mathrm{sgn} \left(n\right) \right\vert,
\label{solu_trivial_anti_1}
\end{equation}
or 
\begin{equation}
\phi _{n}^{\mathrm{A}} = \phi _{n}^{\mathrm{B}} = \sqrt{\frac{\omega - \omega_{0} +J^{\prime}}{\gamma}} 
\mathrm{sgn} \left(n\right),
\label{solu_trivial_anti_2}
\end{equation}
The solutions provided by Eqs. (\ref{solu_trivial_sym_1})-(\ref{solu_trivial_anti_2}) correspond to the dark bulk solitons in the AC limit.

\subsection{Existence of dark bulk solitons and their stability analysis}

Following the same approach, we aim to ensure that the solutions for dark bulk solitons remain consistent with the ideal dark solitons that maintain a constant background, even at the edges of the lattice. To achieve this, we again utilize the nonlinear SSH lattice with appropriately modified boundary conditions, as illustrated in Fig. \ref{fig_structure_bulk}, but with $J$ and $J'$ exchanged.

\begin{figure}[tbp]
\includegraphics[width=8.6cm]{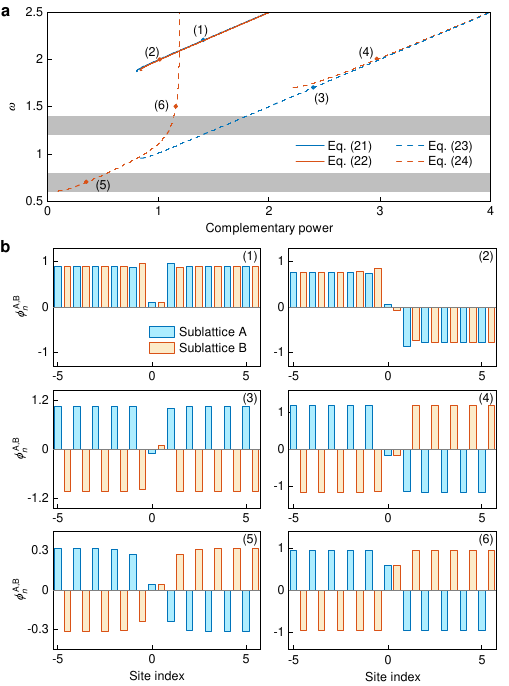}
\caption{\textbf{The existence of dark bulk solitons in the originally topologically trivial lattice.}
\textbf{a} Existence curves of the dark bulk solitons, showing the dependence between frequency and complementary power. The dark bulk solitons, which are continuations of distinct solutions in the AC limit, are labeled in the legend for better understanding. The gray regions denote the frequency bands associated with the bulk states of the originally 
linear SSH lattice.
\textbf{b} Amplitude distributions of the dark bulk solitons, which are marked on the existence curves in (a). 
The blue and red bars represent the amplitudes at sublattice A and sublattice B, respectively. }
\label{fig_trivial_bulk_mode}
\end{figure}

Figure \ref{fig_trivial_bulk_mode}a displays families of different types of dark bulk solitons in the originally topologically trivial SSH lattice with modified boundaries. These dark bulk solitons, which are continuations of distinct solutions in the AC limit, are labeled in the legend for clarity. The frequency bands associated with the linear bulk states are represented by gray regions. The complementary power $P_{r}$, as defined in 
Eq. (\ref{Pr}), is again used to characterize the dark bulk solitons.
Correspondingly, Fig. \ref{fig_trivial_bulk_mode}b shows the amplitude distributions of the dark bulk solitons, 
with the corresponding frequencies indicated on the existence curves in Fig. \ref{fig_trivial_bulk_mode}a.
Note that, for the sake of comparison, we have selected the dark bulk solitons at the same frequencies as those used in 
Fig. \ref{fig_nontrivial_bulk_mode}b.

\begin{figure}[tbp]
\includegraphics[width=8.6cm]{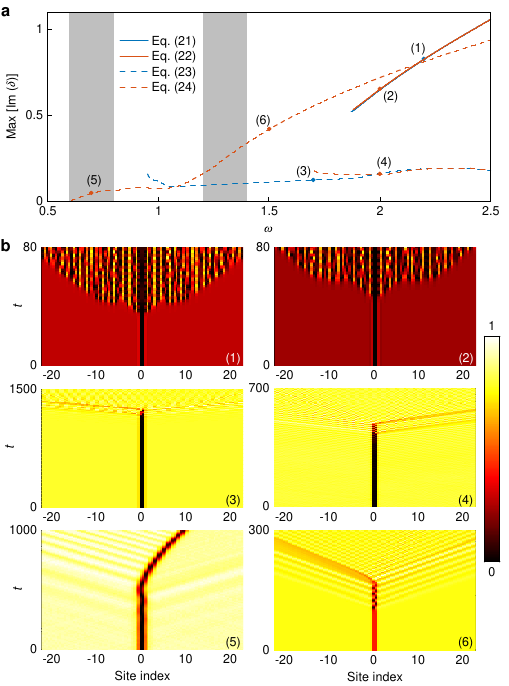}
\caption{\textbf{The stability analysis of dark bulk solitons in the originally topologically trivial lattice.} 
\textbf{a} The maximum growth rates are shown 
for perturbed dark bulk solitons at various frequencies. The dark bulk solitons, which are continuations of distinct solutions in the AC limit, 
are labeled in the legend. The gray regions represent the frequency bands associated with the bulk states of the original linear SSH lattice.
\textbf{b} Temporal evolution of dark bulk solitons selected at specific frequencies indicated on the curves in (a).
For comparison across different cases, the intensities are normalized to their respective maximum values, 
with black representing zero and white representing one.}
\label{fig_trivial_bulk_stability}
\end{figure}

By comparing Fig. \ref{fig_trivial_bulk_mode} with Fig. \ref{fig_nontrivial_bulk_mode}, we find that states (1) and (2) 
in Fig. \ref{fig_trivial_bulk_mode} are very similar to states (1) and (2) in Fig. \ref{fig_nontrivial_bulk_mode}, respectively, in terms of their existence ranges and soliton profiles. Since the soliton amplitudes are equal at the lattice sites away from the center, the amplitude of the constant background can be expressed as
$\phi_{\mathrm{b}} = \pm \sqrt{\left(\omega-\omega_{0}-J-J^{\prime}\right)/\gamma}$.
The key difference between states (1) and (2) in Fig. \ref{fig_trivial_bulk_mode} and states (1) and (2) in Fig. \ref{fig_nontrivial_bulk_mode} 
lies in the coupling characteristics of the lattices. The originally topologically trivial lattice exhibits stronger intracell coupling and weaker intercell coupling, leading to deeper intensity dips for states (1) and (2) in Fig. \ref{fig_trivial_bulk_mode}.
For states (3)-(6) in Fig. \ref{fig_trivial_bulk_mode} and states (3)-(6) in Fig. \ref{fig_nontrivial_bulk_mode}, 
all of them are continuations of the antisymmetric solutions of the single unit cell in the AC limit. Their amplitude 
background can be described by
$\phi_{\mathrm{b}} = \pm \sqrt{\left(\omega-\omega_{0}+J+J^{\prime}\right)/\gamma}.$
However, compared to states (3)-(6) in Fig. \ref{fig_nontrivial_bulk_mode}, states (3)-(6) in Fig. \ref{fig_trivial_bulk_mode} exhibit deeper intensity dips. Furthermore, in contrast to the dark bulk solitons that are continuations of Eq. (\ref{solu_nontrivial_anti_1}), which can only reside in the upper semi-infinite gap, the dark bulk solitons that continue from Eq. (\ref{solu_trivial_anti_1}) can exist not only in the semi-infinite gap but also within the middle finite gap and the top band for the linear bulk states. This behavior supports the previous assertion that, in the case of a topologically nontrivial lattice, an increase in the ratio $J/J^{\prime}$ can allow dark bulk solitons to enter the band of linear bulk states.
Besides, note that for the dark bulk solitons which continue from 
from Eq. (\ref{solu_trivial_anti_2}), one branch exists where the frequency extends from the lower edge of the bottom band of the original linear SSH lattice to positive infinity. This point is the same to the case for the originally topologically
nontrivial lattice.

We then perform a stability analysis of the dark bulk solitons depicted in Fig.\ \ref{fig_trivial_bulk_mode}. Utilizing linear stability analysis, Fig.\ \ref{fig_trivial_bulk_stability}a presents the maximum growth rates, $\mathrm{Max} \left[ \mathrm{Im} \left( \delta \right) \right]$, for the perturbed dark bulk solitons across various frequencies. For reference, the gray regions indicate the frequency bands associated with the bulk states of the original linear SSH lattice. From the figure, we observe that the dark bulk solitons, which continue from Eqs.\ (\ref{solu_trivial_sym_1})-(\ref{solu_trivial_anti_2}), are all linearly unstable. Additionally, we have conducted direct simulations of the temporal evolution of the dark bulk solitons, and the results shown in Fig.\ \ref{fig_trivial_bulk_stability}b align well with those obtained from the linear stability analysis. Note that in the numerical simulations, no extra random perturbations are added to the initial excitations of states (1), (2), (3), and (6), while $\pm 1\%$ random perturbations are introduced to the initial excitations of states (4) and (5) to facilitate the breakup of dark bulk solitons.

\begin{figure}[tbp]
\includegraphics[width=8.6cm]{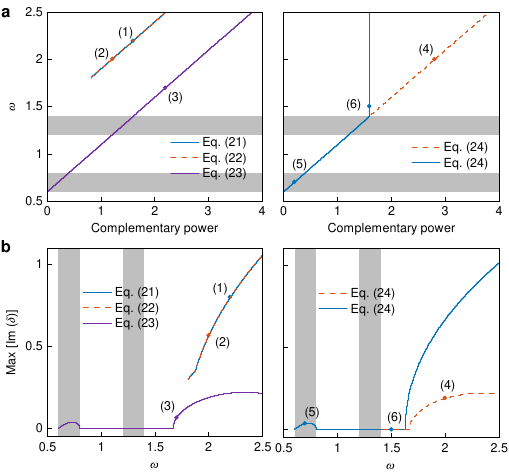}
\caption{\textbf{The existence and stability analysis of dark bulk solitons for $J^{\prime} = 0.399$ and $J = 0.001$.}
\textbf{a} Existence curves of the dark bulk solitons. 
\textbf{b} The maximum growth rates of perturbed dark bulk solitons at various frequencies.
In both (a) and (b), the left and right columns correspond to the dark bulk solitons that originate from 
Eqs. (\ref{solu_trivial_sym_1})-(\ref{solu_trivial_anti_1}) and Eq. (\ref{solu_trivial_anti_2}), respectively.}
\label{fig_trivial_bulk_extreme}
\end{figure}

For this originally topologically trivial lattice, we have recalculated various types of dark bulk solitons for different values of $J^{\prime}$ and $J$, 
ensuring that $J^{\prime} + J = 0.4$ and $J^{\prime}/J > 1$. The latter condition indicates that the originally linear SSH lattice is topologically trivial. 
The dark bulk solitons that originate from Eqs. (\ref{solu_trivial_sym_1})-(\ref{solu_trivial_sym_2}) consistently reside in the semi-infinite gap and are 
always linearly unstable. 
For example, the left columns of Figs. \ref{fig_trivial_bulk_extreme}a and \ref{fig_trivial_bulk_extreme}b show, respectively, 
the existence and stability analysis of these types of dark bulk solitons for $J^{\prime} = 0.399$ and $J = 0.001$.
In contrast, the dark bulk solitons originating from Eq. (\ref{solu_trivial_anti_1}) can enter the bottom band of the linear bulk states as the ratio $J^{\prime}/J$ 
further increases. Specifically, as $J^{\prime}/J$ approaches infinity, these solitons extend from the lower edge of the bottom band of the original linear SSH lattice 
to positive infinity, as indicated by the purple curve in the left column of Fig. \ref{fig_trivial_bulk_extreme}a.
For the dark bulk solitons continuing from Eq. (\ref{solu_trivial_anti_2}), the branch corresponding to state (4) consistently remains in the upper semi-infinite gap. 
In the limit of $J^{\prime}/J = \infty$, the dark bulk solitons in this branch extend from the upper edge of the top band of the original linear SSH lattice to positive infinity, as indicated by the dashed red curve in the right column of Fig. \ref{fig_trivial_bulk_extreme}a.
Under various values of $J^{\prime}$ and $J$, the dark bulk solitons in the branch corresponding to states (5) and (6) occupy the same position as indicated in Fig. \ref{fig_trivial_bulk_mode}a. 
A typical example of this type of dark bulk solitons for $J^{\prime} = 0.399$ and $J = 0.001$ is represented by the solid blue curve in the right column of Fig. \ref{fig_trivial_bulk_extreme}a.
Notably, when the ratio $J^{\prime}/J$ becomes sufficiently large, certain frequency ranges exhibit zero values of $\mathrm{Max} \left[\mathrm{Im}(\delta)\right]$ 
for the dark bulk solitons in the branch corresponding to state (3), the branch corresponding to state (4), and the branch corresponding to states (5) and (6), 
as shown in Fig. \ref{fig_trivial_bulk_extreme}b. This implies that these dark bulk solitons in these frequency ranges are linearly stable. Direct numerical simulations, 
incorporating random perturbations of $\pm 5\%$ added to the initial excitations, demonstrate that the dark bulk solitons within these frequency ranges can 
maintain their intensity dips for at least up to $t = 10^{5}$.

\section{Dark edge solitons in the originally topologically nontrivial lattice}

In this section, we investigate dark edge solitons characterized by intensity dips that are located at the edges of an originally topologically 
nontrivial lattice. Without loss of generality, we focus on the dark edge solitons that are located at the left edge of the nonlinear SSH lattice.

We consider the AC limit with $J^{\prime} = 0$, which implies the absence of intercell coupling. Based on the symmetric solution for a single unit cell, we can construct the solution for the entire lattice as follows:
\begin{equation}
\phi_{n}^{\mathrm{A}} = \phi_{n}^{\mathrm{B}} = \sqrt{\frac{\omega - \omega_{0} - J}{\gamma}} H \left( n-1 \right),
\label{solu_AC_edge_nontrivial_1}
\end{equation}
where $H \left( n \right)$ is the Heaviside function defined as $H \left( n \right)=0$ for $n = 0$ and $H \left( n \right)=1$ for $n \ge 1$. Similarly, based on the antisymmetric solution for a single unit cell, we can construct the solution for the entire lattice as follows:
\begin{equation}
\phi_{n}^{\mathrm{A}} = -\phi_{n}^{\mathrm{B}} = \sqrt{\frac{\omega - \omega_{0} + J}{\gamma}} H \left( n-1 \right).
\label{solu_AC_edge_nontrivial_2}
\end{equation}
It is noteworthy that, the four solutions for dark bulk solitons in the topologically nontrivial lattice, specifically described by 
Eqs. (\ref{solu_nontrivial_sym_1})-(\ref{solu_nontrivial_anti_2}), reduce to two types of solutions for dark edge solitons in the AC limit.

\begin{figure}[tbp]
\includegraphics[width=8.6cm]{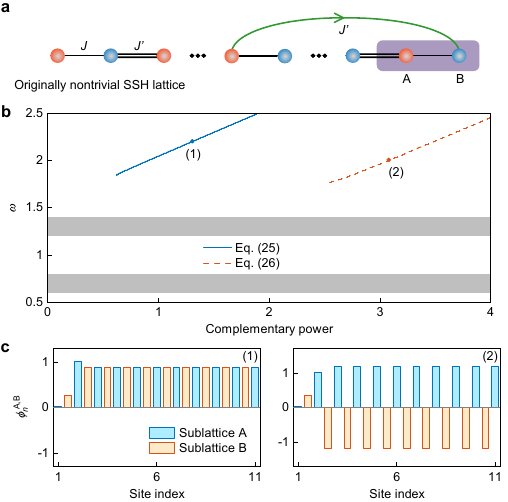}
\caption{\textbf{The existence of dark edge solitons in the originally topologically nontrivial lattice.}  
\textbf{a} Nonlinear SSH lattice where the sublattice site A in the middle unit cell is unidirectionally coupled to the rightmost site 
with the coupling coefficient $J^{\prime}$. This lattice configuration supports dark edge solitons that are continuations of the solutions 
described by Eqs. (\ref{solu_AC_edge_nontrivial_1})-(\ref{solu_AC_edge_nontrivial_2}).  
\textbf{b} Existence curves of the dark edge solitons, illustrating the relationship between frequency and complementary power. 
The dark edge solitons, which continue distinct solutions in the AC limit, are labeled in the legend for clarity. 
The gray regions indicate the frequency bands associated with the bulk states of the original linear SSH lattice.  
\textbf{c} Amplitude distributions for the dark edge solitons, marked on the existence curves in panel (b). 
The blue and red bars represent the amplitudes at sublattice A and sublattice B, respectively. For clarity, 
only soliton amplitudes at several leftmost sites in the lattice are shown.}
\label{fig_nontrivial_edge_mode}
\end{figure}

The solutions provided by Eqs. (\ref{solu_AC_edge_nontrivial_1})-(\ref{solu_AC_edge_nontrivial_2}) correspond to ideal dark solitons positioned at the left edge of the lattice. However, when an open boundary condition is applied directly to the right edge of the lattice, the soliton solutions become perturbed, leading them to deviate from these ideal configurations. To address this issue, we establish a unidirectional coupling between the rightmost lattice site and the sublattice site A in the middle unit cell, characterized by a coupling coefficient $J^{\prime}$, as illustrated in Fig. \ref{fig_nontrivial_edge_mode}a. In condensed matter physics, this unidirectional coupling implies that the right hopping amplitude is $J^{\prime}$, while the left hopping amplitude is zero \cite{PRB103-144202}. This modified boundary condition applies to both types of solutions given by 
Eqs. (\ref{solu_AC_edge_nontrivial_1})-(\ref{solu_AC_edge_nontrivial_2}), as there is no phase jump between the middle and rightmost lattice sites.

Figure \ref{fig_nontrivial_edge_mode}b displays families of different types of dark edge solitons in the originally topologically nontrivial SSH lattice with the modified boundary. The dark edge solitons, which continue distinct solutions in the AC limit, are labeled in the legend for clarity. For comparison, the frequency bands associated with the linear bulk states shown in Figs. \ref{fig_linear}b-c are represented by the gray regions. The complementary power $P_{r}$, as defined in Eq. (\ref{Pr}), characterizes the dark edge solitons. Correspondingly, Fig. \ref{fig_nontrivial_edge_mode}c shows the amplitude distributions of the dark edge solitons, with corresponding frequencies indicated on the existence curves in Fig. \ref{fig_nontrivial_edge_mode}b. Although we only show the soliton amplitudes at several leftmost sites in the lattice, the dark edge solitons illustrated in Fig. \ref{fig_nontrivial_edge_mode}c exhibit constant backgrounds.

\begin{figure}[tbp]
\includegraphics[width=8.6cm]{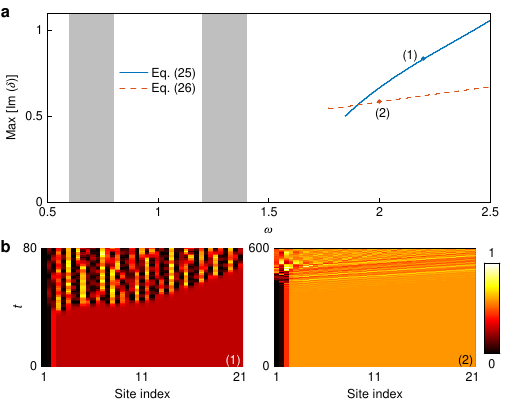}
\caption{\textbf{The stability analysis of dark edge solitons in the originally topologically nontrivial lattice.} 
\textbf{a} Maximum growth rates for perturbed dark edge solitons are presented at various frequencies. The dark edge solitons, which are continuations of distinct solutions in the AC limit, are labeled in the legend. The gray regions represent the frequency bands associated with the bulk states of the original linear SSH lattice.
\textbf{b} Temporal evolution of dark edge solitons selected at specific frequencies indicated on the curves in panel (a). For comparison across different cases, the intensities are normalized to their respective maximum values, with black representing zero and white representing one.}
\label{fig_nontrivial_edge_stability}
\end{figure}

As shown in Fig. \ref{fig_nontrivial_edge_mode}b, the continuations of Eqs. (\ref{solu_AC_edge_nontrivial_1}) and (\ref{solu_AC_edge_nontrivial_2}) yield separate branches of dark edge solitons. Both branches reside spectrally within the semi-infinite gap, and the intensity dips of the dark edge solitons in these branches are well preserved. The existence of these soliton branches can be interpreted as a transition of the dark bulk solitons, illustrated in Fig. \ref{fig_nontrivial_bulk_mode}, from the lattice bulk to the edge. After this transition, since Eqs. (\ref{solu_nontrivial_sym_1}) and (\ref{solu_nontrivial_sym_2}) both reduce to Eq. (\ref{solu_AC_edge_nontrivial_1}), the bulk soliton branches corresponding to states (1) and (2) in Fig. \ref{fig_nontrivial_bulk_mode} evolve into a single edge soliton branch corresponding to state (1) in Figs. \ref{fig_nontrivial_edge_mode}b-c. Similarly, 
since Eqs. (\ref{solu_nontrivial_anti_1}) and (\ref{solu_nontrivial_anti_2}) both reduce to Eq. (\ref{solu_AC_edge_nontrivial_2}), the bulk soliton branches corresponding to states (3) and (4) in Fig. \ref{fig_nontrivial_bulk_mode} evolve into a single edge soliton branch corresponding to state (2) in 
Figs. \ref{fig_nontrivial_edge_mode}b-c. 
In contrast to the dark bulk solitons shown in Fig. \ref{fig_nontrivial_bulk_mode}, dark edge solitons that are continuations of the solution 
given by Eq. (\ref{solu_AC_edge_nontrivial_2}) do not exist when the frequency lies in the middle finite gap. Although we can obtain a family of dark edge solitons by moving the dark bulk solitons from the branch corresponding to states (5) and (6) in Fig. \ref{fig_nontrivial_bulk_mode} to the lattice edge, these dark edge solitons cannot be reduced to the solution given by Eq. (\ref{solu_AC_edge_nontrivial_2}) in the AC limit. Therefore, we exclude this type of dark edge solitons in Figs. \ref{fig_nontrivial_edge_mode}b-c.

We then conduct a stability analysis of the dark edge solitons depicted in Figs.\ \ref{fig_nontrivial_edge_mode}b-c. 
The boundary condition shown in Fig. \ref{fig_nontrivial_edge_mode}a is once again applied here for both the linear stability analysis 
and the dynamical evolutions. Utilizing linear stability analysis, Fig.\ \ref{fig_nontrivial_edge_stability}a presents the maximum growth rates, 
$\mathrm{Max} \left[ \mathrm{Im} \left( \delta \right) \right]$, for the perturbed dark edge solitons across various frequencies. 
The gray regions indicate the frequency bands associated with the bulk states of the original linear SSH lattice.
From the figure, we observe that the dark edge solitons, which continue from Eqs.\ (\ref{solu_AC_edge_nontrivial_1}) and (\ref{solu_AC_edge_nontrivial_2}), are linearly unstable. We have also conducted direct simulations of the temporal evolution of the dark edge solitons without adding any extra random perturbations to the initial excitations of states (1) and (2). The results shown in Fig.\ \ref{fig_nontrivial_edge_stability}b reveal a breakup of intensity during the temporal evolution, indicating that the dark edge solitons are dynamically unstable. This finding is consistent with the conclusions drawn from the linear stability analysis.

In addition, we have calculated various types of dark edge solitons for different values of $J$ and $J'$, ensuring that $J + J' = 0.4$ and $J/J' < 1$, which maintains the original SSH lattice as topologically nontrivial.
The dark edge solitons derived from Eqs. (\ref{solu_AC_edge_nontrivial_1})-(\ref{solu_AC_edge_nontrivial_2}) consistently reside in the semi-infinite gap and are always linearly unstable. Meanwhile, we have not found any dark edge solitons residing at the middle frequency
$\omega =1$, even when $J = J^{\prime}$. This point will be further elaborated upon in the following study of dark edge solitons in the originally topologically trivial lattice.

\section{Dark edge solitons in the originally topologically trivial lattice}

\begin{figure}[tbp]
\includegraphics[width=8.6cm]{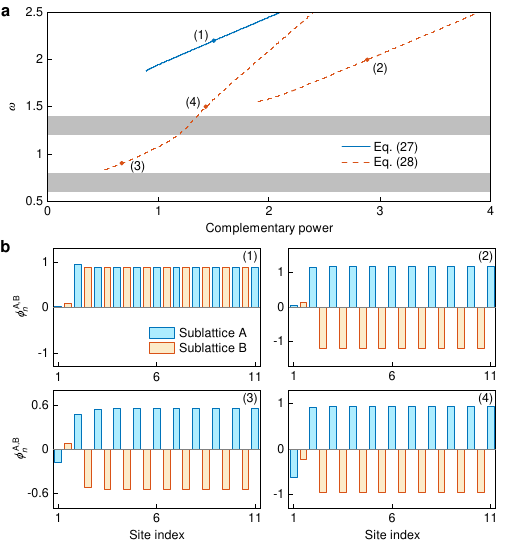}
\caption{\textbf{The existence of dark edge solitons in an originally topologically trivial lattice.}  
\textbf{a} Existence curves of the dark edge solitons, illustrating the relationship between frequency and complementary power. The dark edge solitons, which correspond to distinct solutions in the AC limit, are labeled in the legend for clarity. The gray regions indicate the frequency bands associated with the bulk states of the original linear SSH lattice.  
\textbf{b} Amplitude distributions for the dark edge solitons, indicated on the existence curves in panel (a). The blue and red bars represent the amplitudes at sublattice A and sublattice B, respectively.}
\label{fig_trivial_edge_mode}
\end{figure}

To search for dark edge solitons in the originally topologically trivial lattice, we consider the AC limit with $J = 0$. By substituting $J$ in Eq. (\ref{solu_AC_edge_nontrivial_1}) with $J^{\prime}$, we can derive the following solution:
\begin{equation}
\phi_{n}^{\mathrm{A}} = \phi_{n}^{\mathrm{B}} = \sqrt{\frac{\omega - \omega_{0} - J^{\prime}}{\gamma}} H \left( n-1 \right).
\label{solu_AC_edge_trivial_1}
\end{equation}
This solution is based on the symmetric solution for a single unit cell. Additionally, we obtain another solution based on the antisymmetric solution for a single unit cell:
\begin{equation}
\phi_{n}^{\mathrm{A}} = -\phi_{n}^{\mathrm{B}} = \sqrt{\frac{\omega - \omega_{0} + J^{\prime}}{\gamma}} H \left( n-1 \right).
\label{solu_AC_edge_trivial_2}
\end{equation}
Thus, the dark edge solitons can then be obtained based on these solutions. Following the same methodology, to ensure that the solutions for dark edge solitons remain consistent with the ideal dark solitons that maintain a constant background, we again utilize the nonlinear SSH lattice with appropriately modified boundary conditions, as illustrated in Fig.\ \ref{fig_nontrivial_edge_mode}a, but with $J$ and $J'$ exchanged.

Figure \ref{fig_trivial_edge_mode}a displays families of different types of dark edge solitons in the originally topologically trivial SSH lattice with a modified boundary. The dark edge solitons, which correspond to distinct solutions in the AC limit, are labeled in the legend for clarity. For comparison, the frequency bands associated with the linear bulk states, as shown in Figs. \ref{fig_linear}e-f, are represented by the gray regions. The complementary power $P_{r}$, as defined in Eq. (\ref{Pr}), characterizes the dark edge solitons. Correspondingly, Fig. \ref{fig_trivial_edge_mode}b illustrates the amplitude distributions of the dark edge solitons, with the corresponding frequencies indicated on the existence curves in Fig. \ref{fig_trivial_edge_mode}a. Again, although we only show the soliton amplitudes at several leftmost sites in the lattice, the dark edge solitons illustrated in Fig. \ref{fig_trivial_edge_mode}b exhibit constant backgrounds.

\begin{figure}[tbp]
\includegraphics[width=8.6cm]{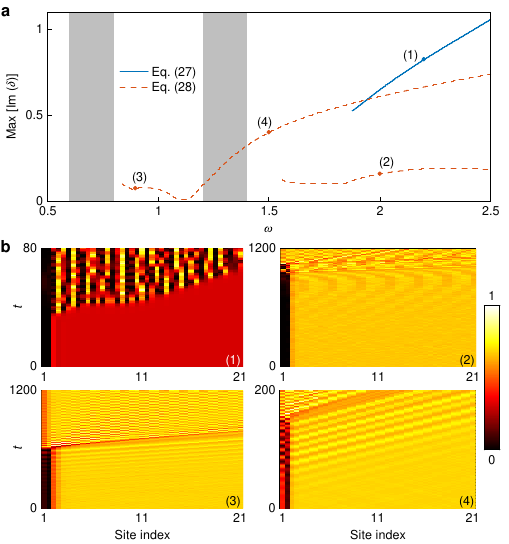}
\caption{\textbf{The stability analysis of dark edge solitons in the originally topologically trivial lattice.}
\textbf{a} Maximum growth rates for perturbed dark edge solitons are shown for various frequencies. 
The dark edge solitons, which are continuations of distinct solutions in the AC limit, are labeled in the legend. 
The gray regions indicate the frequency bands associated with the bulk states of the original linear SSH lattice.
\textbf{b} Temporal evolution of dark edge solitons selected at specific frequencies indicated in panel (a). 
For comparison across different cases, the intensities are normalized to their respective maximum values, where black represents zero and white represents one.}
\label{fig_trivial_edge_stability}
\end{figure}

As shown in Fig. \ref{fig_trivial_edge_mode}a, the continuations of Eqs. (\ref{solu_AC_edge_trivial_1}) and (\ref{solu_AC_edge_trivial_2}) yield three branches of dark edge solitons. The existence of these soliton branches can be interpreted as a transition of the dark bulk solitons, illustrated in Fig. \ref{fig_trivial_bulk_mode}, from the bulk of the lattice to the edge. After this transition, since Eqs. (\ref{solu_trivial_sym_1}) and (\ref{solu_trivial_sym_2}) both reduce to Eq. (\ref{solu_AC_edge_trivial_1}), the bulk soliton branches corresponding to states (1) and (2) in Fig. \ref{fig_trivial_bulk_mode} evolve into a single edge soliton branch corresponding to state (1) in Fig. \ref{fig_trivial_edge_mode}. Similarly, since Eqs. (\ref{solu_trivial_anti_1}) and (\ref{solu_trivial_anti_2}) both reduce to Eq. (\ref{solu_AC_edge_trivial_2}), the segment of the bulk soliton branch corresponding to state (3), which resides in the semi-infinite gap, and the entire bulk soliton branch corresponding to state (4), both shown in Fig. \ref{fig_trivial_bulk_mode}, evolve into a single edge soliton branch corresponding to state (2) in Fig. \ref{fig_trivial_edge_mode}.
Additionally, unlike the case shown in Fig. \ref{fig_nontrivial_edge_mode}, the remaining segment of the bulk soliton branch corresponding to state (3) and the bulk soliton branch corresponding to states (5) and (6) in Fig. \ref{fig_trivial_bulk_mode} evolve into another edge soliton branch corresponding to states (3) and (4) in Fig. \ref{fig_trivial_edge_mode}. We have confirmed that the dark edge solitons with $\omega < \omega_{0} + J'$ in this branch can be reduced to the solutions in the AC limit, specifically Eq. (\ref{solu_AC_edge_trivial_2}). For this reason, we have included this branch in Fig. \ref{fig_trivial_edge_mode}. Furthermore, the intensity dips of all dark edge solitons in the three branches are always well preserved, remaining unaffected by the band structure of the originally linear lattice.

We then conduct a stability analysis of the dark edge solitons depicted in Fig. \ref{fig_trivial_edge_mode}. Utilizing linear stability analysis, Fig. \ref{fig_trivial_edge_stability}a presents the maximum growth rates, $\mathrm{Max} \left[ \mathrm{Im} \left( \delta \right) \right]$, for the perturbed dark edge solitons across various frequencies. The gray regions indicate the frequency bands associated with the bulk states of the original linear SSH lattice. From the figure, since $\mathrm{Max} \left[ \mathrm{Im} \left( \delta \right) \right] > 0$ for all the curves, we conclude that all dark edge solitons, including those residing in the middle finite gap, are linearly unstable. 
We have also performed direct simulations of the temporal evolution of these dark edge solitons, with the results shown in Fig. \ref{fig_trivial_edge_stability}b. The observed breakup of intensity during the temporal evolution implies that the dark edge solitons are dynamically unstable, which is consistent with the conclusions drawn from the linear stability analysis. Notably, to expedite the breakup of states (2)-(4) shown in Fig. \ref{fig_trivial_edge_mode}, we introduced $\pm 1\%$ random perturbations to the initial input, while no additional perturbations were applied for the initial excitation of state (1).

\begin{figure}[tbp]
\includegraphics[width=8.6cm]{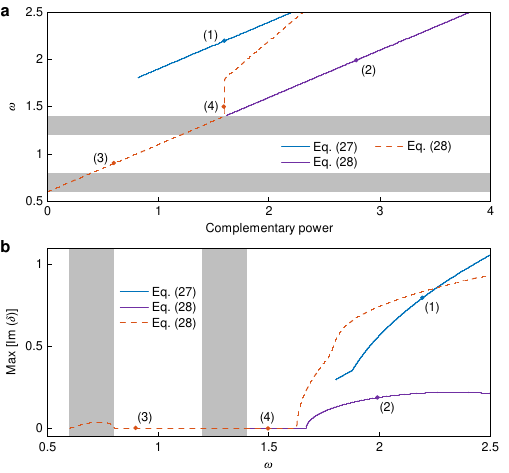}
\caption{\textbf{The existence and stability analysis of dark edge solitons for $J^{\prime} = 0.399$ and $J = 0.001$.}
\textbf{a} Existence curves of the dark edge solitons. 
\textbf{b} The maximum growth rates of perturbed dark edge solitons at various frequencies.
In both (a) and (b), the dark edge solitons, which are continuations of distinct solutions in the AC limit, are labeled in the legend.}
\label{fig_trivial_edge_extreme}
\end{figure}

In addition, we have recalculated various types of dark edge solitons for different values of $J^{\prime}$ and $J$, 
ensuring that $J^{\prime} + J = 0.4$ and $J^{\prime}/J > 1$. The latter condition preserves the original SSH lattice as topologically trivial. 
The dark edge solitons that continue from Eq. (\ref{solu_AC_edge_trivial_1}) consistently reside in the semi-infinite gap and are always linearly unstable. 
For example, Figs. \ref{fig_trivial_edge_extreme}a and \ref{fig_trivial_edge_extreme}b show, respectively, 
the existence and stability analysis of this type of dark edge solitons for $J^{\prime} = 0.399$ and $J = 0.001$.
For the dark edge solitons continuing from Eq. (\ref{solu_AC_edge_trivial_2}), the branch corresponding to state (2) remains in the upper semi-infinite gap.
In the limit of $J^{\prime}/J = \infty$, the dark edge solitons in this branch extend from the upper edge of the top band of the original linear SSH lattice to positive infinity, as indicated by the purple curve in Fig. \ref{fig_trivial_edge_extreme}a.
The dark edge solitons in the branch corresponding to states (3) and (4) can enter the bottom band of the linear bulk states as the ratio $J^{\prime}/J$ further increases. 
Specifically, as $J^{\prime}/J$ approaches infinity, these solitons extend from the lower edge of the bottom band of the original linear SSH lattice to positive infinity, as indicated by the dashed red curve in Fig. \ref{fig_trivial_edge_extreme}a.
Notably, when the ratio $J^{\prime}/J$ becomes sufficiently large, certain frequency ranges exhibit zero values of $\mathrm{Max} \left[\mathrm{Im}(\delta)\right]$ 
for the dark edge solitons in the branch corresponding to state (2), and the branch corresponding to states (3) and (4),
as shown in Fig. \ref{fig_trivial_edge_extreme}b. This indicates that these dark edge solitons within these frequency ranges are linearly stable. 
Direct numerical simulations, which incorporate random perturbations of $\pm 5\%$ added to the initial excitations, 
demonstrate that the dark edge solitons in these frequency ranges can maintain their intensity dips for at least up to $t = 10^{5}$.

\section{Discussion}

In this section, we discuss the differences between our work and previous studies on dark solitons. In conventional nonlinear lattices, 
the concept of topological phases has not been introduced, leading most studies to focus primarily on the existence and stability analysis of 
various types of dark solitons within different lattice structures \cite{RMP83-247}. The relative spectral position of dark solitons 
concerning the band structure of the original linear lattice, as well as changes in their properties across parameter regimes corresponding to different topological phases, are seldom addressed. In contrast, in our work, we demonstrate that in nonlinear SSH lattices, dark solitons with intensity dips located either in the bulk of the lattice or at its edges can reside within the semi-infinite gap or the middle finite gap. Furthermore, through a systematic study of dark solitons across a wide range of parameters, we find that several types of dark solitons exhibit linear stability when the intracell coupling is significantly larger than the intercell coupling, which corresponds to the system being deep inside the topologically trivial phase.

When the concept of topological phases is introduced into nonlinear lattices, specifically in nonlinear topological lattices, dark topological edge solitons have been proposed in both two-dimensional and three-dimensional settings \cite{PRA97-043602,nano10-3559,CSF186-115239,NJP22-103058}. These dark topological edge solitons arise from the balance between dispersion and nonlinearity along the propagation direction of the linear topological edge states. Specifically, they can be regarded as hybrid states, where the transverse direction is characterized by linear topological edge states, while the longitudinal propagation direction supports dark soliton solutions in free space. However, in our one-dimensional lattices, there is only one spatial direction, with nonlinearity acting purely in the transverse direction under the modulation of staggered hopping amplitudes. Unlike previous dark topological edge solitons that exclusively reside within the topological bandgap, the intensity dip of our dark solitons remains unaffected by the band structure of the original linear lattices and is always well preserved, regardless of the specific soliton type.

Finally, we would like to note that the dark solitons we have discovered may not have a clear and direct connection to topological phases, unlike the topological edge states or topological edge solitons. Nevertheless, we find that the type, existence, and stability of dark solitons exhibit different properties in the topologically nontrivial and trivial phases. In particular, stable dark solitons are found only in a parameter regime where the system is deep within the topologically trivial phase. These results provide insight into soliton formation and properties in nonlinear topological lattices.

\section{Conclusion}

In conclusion, we have systematically investigated dark solitons in nonlinear SSH lattices. For dark bulk solitons in originally topologically nontrivial lattices, 
they may spectrally reside in the semi-infinite gap or the middle finite gap, but they are always dynamically unstable. In contrast, dark bulk solitons in originally 
topologically trivial lattices can also exist in the semi-infinite gap or the middle finite gap. Notably, when the intracell coupling is significantly larger than the intercell coupling, 
dark bulk solitons become linearly stable in certain frequency ranges.
Regarding dark edge solitons in originally topologically nontrivial lattices, they can only spectrally reside in the semi-infinite gap and are always dynamically unstable. 
Conversely, dark edge solitons in originally topologically trivial lattices may reside in the semi-infinite gap or the middle finite gap, 
with specific regimes for linear stability emerging when intracell coupling is much larger than intercell coupling.
Importantly, for all types of dark solitons, the intensity dips are consistently well preserved, remaining unaffected by the band structure of the original linear SSH lattice. 
Regarding the experimental realization of nonlinear SSH lattices, electric circuit lattices emerge as a promising platform for implementing the long-range couplings 
depicted in Figs. \ref{fig_structure_bulk} and \ref{fig_nontrivial_edge_mode}a, especially when the couplings are negative or unidirectional \cite{NSR8-nwaa192,CP8-342,nelectron1-178,
PRL133-116602,FP18-33311,CSF157-111912,FP20-44204}. The requirement for long-range couplings may be relaxed when the lattice size is sufficiently large, 
allowing us to neglect the perturbations induced by the lattice edges on the constant soliton background. In this context, the dark solitons we propose 
may be observed in photonic waveguide arrays \cite{OL45-6466,PRL128-093901,LSA9-147,science372-72,nanophotonics14-769}, 
polaritonic systems \cite{nphys18-678}, or Bose-Einstein condensates \cite{PRB108-184301,LSA14-296}.
These findings may provide valuable insights for exploring novel types of solitons in nonlinear topological lattices.

\section*{Acknowledgements}

R.L. and W.W. were sponsored by the National Key Research and Development Program
of China (Grant No. 2022YFA1404902), National Natural Science Foundation
of China (Grant No. 12104353), and Fundamental Research Funds for the Central
Universities (Grant No. QTZX25086). Y.L. was sponsored by the National Natural
Science Foundation of China (NSFC) under Grant No. 62271366 and the 111 Project.
Numerical calculations performed in this work were supported by the High-
Performance Computing Platform of Xidian University.

\section*{Data availability}

The data that support the findings of this article are not
publicly available upon publication because it is not technically
feasible and/or the cost of preparing, depositing, and
hosting the data would be prohibitive within the terms of this
research project. The data are available from the authors upon
reasonable request.


\begin{thebibliography}{99}

\bibitem{TI1} C. L. Kane and E. J. Mele, Quantum Spin Hall Effect in Graphene,
Phys. Rev. Lett. 95, 226801 (2005).

\bibitem{TI2} M. K\"{o}nig, S. Wiedmann, C. Brüne, A. Roth, H. Buhmann, L. W.
Molenkamp, X.-L. Qi, and S.-C. Zhang, Quantum Spin Hall Insulator State in HgTe
Quantum Wells, Science 318, 766 (2007).

\bibitem{RMP82-3045} M. Z. Hasan and C. L. Kane, Colloquium: Topological insulators,
Rev. Mod. Phys. 82, 3045 (2010).

\bibitem{RMP83-1057} X.-L. Qi and S.-C. Zhang, Topological insulators and superconductors,
Rev. Mod. Phys. 83, 1057 (2011).

\bibitem{QF3-14} F. Tang and X. Wan, Group-theoretical study of band nodes and the emanating nodal structures in
crystalline materials, Quantum Front. 3, 14 (2024).

\bibitem{QF2-22} G.-Q. Zhao, S. Li, W. B. Rui, C. M. Wang, H.-Z. Lu, and X. C. Xie, 3D quantum Hall effect in a topological 
nodal-ring semimetal, Quantum Front. 2, 22 (2023).

\bibitem{QF3-21} F. Zhan, R. Chen, Z. Ning, D.-S. Ma, Z. Wang, D.-H. Xu, and R. Wang, Perspective: Floquet engineering topological
states from effective models towards realistic materials, Quantum Front. 3, 21 (2024).

\bibitem{NRP1-281} G. Ma, M. Xiao, and C. T. Chan, Topological phases in acoustic and
mechanical systems, Nat. Rev. Phys. 1, 281 (2019).

\bibitem{NRM7-974} H. Xue, Y. Yang, and B. Zhang, Topological acoustics, Nat. Rev. Mater. 7, 974 (2022).

\bibitem{NRP5-483} Z.-K. Lin, Q. Wang, Y. Liu, H. Xue, B. Zhang, Y. Chong, and J.-H. Jiang,
Topological phenomena at defects in acoustic, photonic and solid-state lattices, Nat. Rev. Phys. 5, 483 (2023).

\bibitem{RMP96-021002} T. Shah, C. Brendel, V. Peano, and F. Marquardt, Colloquium: Topologically
protected transport in engineered mechanical systems, Rev. Mod. Phys. 96, 021002 (2024).

\bibitem{QF3-26} Y. Liu, K. Li, W. Liu, Z. Zhang, Y. Cheng, and X. Liu, Observation of chiral Landau levels in two-dimensional
acoustic system, Quantum Front. 3, 26 (2024).

\bibitem{RPP86-106501} W. Zhu, W. Deng, Y. Liu, J. Lu, H.-X. Wang, Z.-K. Lin, X. Huang, J.-H. Jiang, and Z. Liu, 
Topological phononic metamaterials, Rep. Prog. Phys. 86, 106501 (2023).

\bibitem{RMP91-015005} N. R. Cooper, J. Dalibard, and I. B. Spielman, Topological bands for ultracold
atoms, Rev. Mod. Phys. 91, 015005 (2019).

\bibitem{nphys12-639} N. Goldman, J. C. Budich, and P. Zoller, Topological quantum matter with ultracold 
gases in optical lattices, Nature Phys. 12, 639 (2016).

\bibitem{Segev} M. C. Rechtsman, J. M. Zeuner, Y. Plotnik, Y. Lumer, D. Podolsky, F. Dreisow, S. Nolte,
M. Segev, and A. Szameit, Photonic Floquet topological insulators, Nature 496, 196 (2013).

\bibitem{RMP91-015006} T. Ozawa, H. M. Price, A. Amo, N. Goldman, M. Hafezi, L. Lu, M. C. Rechtsman,
D. Schuster, J. Simon, O. Zilberberg, and I. Carusotto, Topological Photonics, Rev. Mod. Phys. 91, 015006 (2019).

\bibitem{LSA9-130} M. Kim, Z. Jacob, and J. Rho, Recent advances in 2D, 3D and higher-order topological
photonics, Light: Science \& Applications 9, 1 (2020).

\bibitem{QF1-10} M. Yang, J.-S. Xu, C.-F. Li, and G.-C. Guo, Simulating topological materials with photonic synthetic
dimensions in cavities, Quantum Front. 1, 10 (2022).

\bibitem{PIER} J.-W. Liu, G.-G. Liu, and B. Zhang, Three-dimensional Topological Photonic Crystals (Invited Review), 
PIER 181, 99 (2024).

\bibitem{APR} Z. Guo, Y. Wang, S. Ke, X. Su, J. Ren, and H. Chen, 1D Photonic Topological Insulators Composed of 
Split Ring Resonators: A Mini Review, Advanced Physics Research 2300125 (2023).

\bibitem{PR1093-1} H. Yang, L. Song, Y. Cao, and P. Yan, Circuit realization of topological physics, 
Physics Reports 1093, 1 (2024).

\bibitem{APLED} H. Sahin, M. B. A. Jalil, and C. H. Lee, Topolectrical circuits---Recent experimental advances 
and developments, APL Electronic Devices 1, 021503 (2025).

\bibitem{NSR8-nwaa192} R. Li, B. Lv, H. Tao, J. Shi, Y. Chong, B. Zhang, and H. Chen, 
Ideal type-II Weyl points in topological circuits, National Science Review 8, nwaa192 (2021).

\bibitem{PRR2-023180} S. A. Hassani Gangaraj, C. Valagiannopoulos, and F. Monticone, Topological scattering resonances at ultralow frequencies, Phys. Rev. Research 2, 023180 (2020).


\bibitem{APR7-021306} D. Smirnova, D. Leykam, Y. Chong, and Y. Kivshar, Nonlinear topological photonics, 
Appl. Phys. Rev. 7, 021306 (2020).

\bibitem{NP20-905} A. Szameit and M. C. Rechtsman, Discrete nonlinear topological photonics, Nat. Phys. 20, 905 (2024).

\bibitem{PRA90-023813} M. J. Ablowitz, C. W. Curtis, and Y.-P. Ma, Linear and nonlinear traveling edge waves in optical 
honeycomb lattices, Phys. Rev. A 90, 023813 (2014).

\bibitem{PRA94-021801} Y. Lumer, M. C. Rechtsman, Y. Plotnik, and M. Segev, Instability of bosonic topological edge states in the presence 
of interactions, Phys. Rev. A 94, 021801 (2016).

\bibitem{optica3-1228} Y. V. Kartashov and D. V. Skryabin, Modulational instability and solitary waves in polariton topological insulators, 
Optica 3, 1228 (2016).

\bibitem{PRL119-253904} Y. V. Kartashov and D. V. Skryabin, Bistable Topological Insulator with Exciton-Polaritons, 
Phys. Rev. Lett. 119, 253904 (2017).

\bibitem{PRB102-115411} T. Tuloup, R. W. Bomantara, C. H. Lee, and J. Gong, Nonlinearity induced topological physics in 
momentum space and real space, Phys. Rev. B 102, 115411 (2020).

\bibitem{OL45-6466} M. Guo, S. Xia, N. Wang, D. Song, Z. Chen, and J. Yang, Weakly nonlinear topological gap solitons in 
Su-Schrieffer-Heeger photonic lattices, Opt. Lett. 45, 6466 (2020).

\bibitem{nphys18-678} N. Pernet, P. St-Jean, D. D. Solnyshkov, G. Malpuech, N. C. Zambon, Q. Fontaine, 
B. Real, O. Jamadi, A. Lemaître, M. Morassi, L. L. Gratiet, T. Baptiste, A. Harouri, I. Sagnes, A. Amo, S. Ravets, 
and J, Bloch, Gap solitons in a one-dimensional driven-dissipative topological lattice, Nat. Phys. 18, 678 (2022).

\bibitem{PRL128-093901} Y. V. Kartashov et al., Observation of Edge Solitons in Topological Trimer Arrays, 
Phys. Rev. Lett. 128, 093901 (2022).

\bibitem{PRE104-054206} Y.-P. Ma and H. Susanto, Topological edge solitons and their stability in a nonlinear 
Su-Schrieffer-Heeger model, Phys. Rev. E 104, 054206 (2021).

\bibitem{CP8-342} R. Li, X. Kong, W. Wang, Y. Wang, Y. Jia, H. Tao, P. Li, Y. Liu, and B. A. Malomed, 
Observation of edge solitons and transitions between them in a trimer circuit lattice, Commun. Phys. 8, 342 (2025).

\bibitem{PRL117-143901} D. Leykam and Y. D. Chong, Edge Solitons in Nonlinear-Photonic Topological Insulators, 
Phys. Rev. Lett. 117, 143901 (2016).

\bibitem{ACSPhoton7-735} S. K. Ivanov, Y. V. Kartashov, A. Szameit, L. Torner, and V. V. Konotop, 
Vector Topological Edge Solitons in Floquet Insulators, ACS Photonics 7, 735 (2020).

\bibitem{ncommun11-1902} Z. Zhang, R. Wang, Y. Zhang, Y. V. Kartashov, F. Li, H. Zhong, H. Guan, K. Gao, F. Li, 
Y. Zhang, and M. Xiao, Observation of edge solitons in photonic graphene, Nat. Commun. 11, 1902 (2020).

\bibitem{PRX11-041057} S. Mukherjee and M. C. Rechtsman, Observation of Unidirectional Solitonlike Edge States in Nonlinear Floquet Topological Insulators, Phys. Rev. X 11, 041057 (2021).

\bibitem{PRA103-053507} S. K. Ivanov, Y. V. Kartashov, M. Heinrich, A. Szameit, L. Torner, and V. V. Konotop, Topological dipole Floquet solitons, Phys. Rev. A 103, 053507 (2021).

\bibitem{ACSPhoton8-1077} Z. Shi, M. Zuo, H. Li, D. Preece, Y. Zhang, and Z. Chen, Topological Edge States and Solitons on a Dynamically Tunable Domain Wall of Two Opposing Helical Waveguide Arrays, ACS Photonics 8, 1077 (2021).

\bibitem{PRB106-195423} M. Ezawa, Nonlinearity-induced chiral solitonlike edge states in Chern systems, Phys. Rev. B 106, 195423 (2022).

\bibitem{nphys17-995} M. S. Kirsch, Y. Zhang, M. Kremer, L. J. Maczewsky, S. K. Ivanov, Y. V. Kartashov, L. Torner, 
D. Bauer, A. Szameit, and M. Heinrich, Nonlinear second-order photonic topological insulators, Nat. Phys. 17, 995 (2021).

\bibitem{PRL111-243905} Y. Lumer, Y. Plotnik, M. C. Rechtsman, and M. Segev, Self-Localized States in 
Photonic Topological Insulators, Phys. Rev. Lett. 111, 243905 (2013).

\bibitem{PRL118-023901} D. D. Solnyshkov, O. Bleu, B. Teklu, and G. Malpuech, Chirality of Topological Gap 
Solitons in Bosonic Dimer Chains, Phys. Rev. Lett. 118, 023901 (2017).

\bibitem{PRA98-013827} A. N. Poddubny and D. A. Smirnova, Ring Dirac solitons in nonlinear topological systems, Phys. Rev. A 98, 013827 (2018).

\bibitem{LPR13-1900223} D. A. Smirnova, L. A. Smirnov, D. Leykam, and Y. S. Kivshar, Topological Edge States and Gap Solitons in the Nonlinear Dirac Model, Laser \& Photonics Reviews 13, 1900223 (2019).

\bibitem{arxiv1904-10312} J. L. Marzuola, M. Rechtsman, B. Osting, and M. Bandres, Bulk soliton dynamics in bosonic topological insulators, arXiv:1904.10312  (2019).

\bibitem{science368-856} S. Mukherjee and M. C. Rechtsman, Observation of Floquet solitons in a topological bandgap, Science 368, 856 (2020).

\bibitem{CP5-275} R. Li, X. Kong, D. Hang, G. Li, H. Hu, H. Zhou, Y. Jia, P. Li, and Y. Liu, Topological bulk solitons in a nonlinear photonic Chern insulator, Commun. Phys. 5, 275 (2022).

\bibitem{PRL123- 053902} F. Zangeneh-Nejad and R. Fleury, Nonlinear Second-Order Topological Insulators, Phys. Rev. Lett. 123, 053902 (2019).

\bibitem{PRResearch5-L012041} H. Hohmann, T. Hofmann, T. Helbig, S. Imhof, H. Brand,
L. K. Upreti, A. Stegmaier, A. Fritzsche, T. M\"{u}ller, U. Schwingenschl\"{o}gl, C. H. Lee,
M. Greiter, L. W. Molenkamp, T. Kie\ss ling, and R.Thomale, Observation of Cnoidal Wave
Localization in Nonlinear Topolectric Circuits, Phys. Rev. Research 5, L012041 (2023).

\bibitem{PRB93-15512} Y. Hadad, A. B. Khanikaev, and A. Al\`{u}, Self-induced topological transitions 
and edge states supported by nonlinear staggered potentials, Phys. Rev. B 93, 155112 (2016).

\bibitem{nelectron1-178} Y. Hadad, J. C. Soric, A. B. Khanikaev, and A. Al\`{u}, Self-induced topological protection in nonlinear circuit arrays, Nat. Electron. 1, 178 (2018).

\bibitem{ncommun10-1102} Y. Wang, L.-J. Lang, C. H. Lee, B. Zhang, and Y. D. Chong, Topologically enhanced harmonic generation in a nonlinear transmission line metamaterial, Nat. Commun. 10, 1102 (2019).

\bibitem{nnano14-2} S. Kruk, A. Poddubny, D. Smirnova, L. Wang, A. Slobozhanyuk, A. Shorokhov, I. Kravchenko, 
B. Luther-Davies, and Y. Kivshar, Nonlinear light generation in topological nanostructures, Nature Nanotech. 14, 2 (2019).

\bibitem{PRA97-043602} C. Shang, Y. Zheng, and B. A. Malomed, Weyl solitons in three-dimensional optical lattices, 
Phys. Rev. A 97, 043602 (2018).

\bibitem{nano10-3559} B. Ren, H. Wang, V. O. Kompanets, Y. V. Kartashov, Y. Li, and Y. Zhang, Dark topological 
valley Hall edge solitons, Nanophotonics 10, 3559 (2021).

\bibitem{CSF186-115239} S. K. Ivanov and Y. V. Kartashov, Floquet valley Hall edge solitons, Chaos, 
Solitons \& Fractals 186, 115239 (2024).

\bibitem{NJP22-103058} Y.-L. Tao, N. Dai, Y.-B. Yang, Q.-B. Zeng, and Y. Xu, Hinge solitons in three-dimensional 
second-order topological insulators, New J. Phys. 22, 103058 (2020).

\bibitem{arxiv-prb} R. Li, W. Wang, X. Kong, C. Shang, Y. Jia, G.-G. Liu, Y. Liu, and B. Zhang,
Self-Induced Topological Edge States in a Lattice with Onsite Nonlinearity, arXiv:2504.11964.

\bibitem{CP8-451} C. Huang, A. V. Kireev, Y. Jiang, V. O. Kompanets, C. Shang, Y. V. Kartashov, S. A. Zhuravitskii, N. N. Skryabin, 
I. V. Dyakonov, A. A. Kalinkin, S. P. Kulik, F. Ye, and V. N. Zadkov, Observation of nonlinear higher-order topological insulators with unconventional boundary truncations, Commun. Phys. 8, 451 (2025).

\bibitem{CSF207-118044} R. Li, W. Wang, Y. Jia, Y. Liu, P. Li, and B. A. Malomed, Nonlinear quadrupole topological insulators, 
Chaos, Solitons and Fractals 207, 118044 (2026).


\bibitem{PRL121-163901} D. A. Dobrykh, A. V. Yulin, A. P. Slobozhanyuk, A. N. Poddubny, and Yu. S. Kivshar, 
Nonlinear Control of Electromagnetic Topological Edge States, Phys. Rev. Lett. 121, 163901 (2018).

\bibitem{PRB104-235420} M. Ezawa, Nonlinearity-induced transition in the nonlinear Su-Schrieffer-Heeger model and 
a nonlinear higher-order topological system, Phys. Rev. B 104, 235420 (2021).

\bibitem{LSA9-147} S. Xia, D. Juki\'{c}, N. Wang, D. Smirnova, L. Smirnov, L. Tang, D. Song, A. Szameit, D. Leykam, 
J. Xu, Z. Chen, and H. Buljan, Nontrivial coupling of light into a defect: the interplay of nonlinearity and topology, 
Light Sci. Appl. 9, 147 (2020).

\bibitem{science372-72} S. Xia, D. Kaltsas, D. Song, I. Komis, J. Xu, A. Szameit, H. Buljan, K. G. Makris, and Z. Chen, 
Nonlinear tuning of PT symmetry and non-Hermitian topological states, Science 372, 72 (2021).

\bibitem{LSA10-164} Z. Hu, D. Bongiovanni, D. Juki\'{c}, E. Jajti\'{c}, S. Xia, D. Song, J. Xu, R. Morandotti, 
H. Buljan, and Z. Chen, Nonlinear control of photonic higher-order topological bound states in the continuum, 
Light Sci. Appl. 10, 164 (2021).

\bibitem{nanophotonics14-769} C. J\"{o}rg, M. J\"{u}rgensen, S. Mukherjee, and M. C. Rechtsman, 
Optical control of topological end states via soliton formation in a 1D lattice, Nanophotonics 14, 769 (2025).

\bibitem{PRL127-184101} D. Bongiovanni, D. Juki\'{c}, Z. Hu, F. Luni\'{c}, Y. Hu, D. Song, R. Morandotti, Z. Chen, and H. Buljan, 
Dynamically Emerging Topological Phase Transitions in Nonlinear Interacting Soliton Lattices, Phys. Rev. Lett. 127, 184101 (2021).

\bibitem{PRL133-116602} K. Bai, J.-Z. Li, T.-R. Liu, L. Fang, D. Wan, and M. Xiao, Arbitrarily Configurable Nonlinear Topological Modes, Phys. Rev. Lett. 133, 116602 (2024).

\bibitem{arXiv:2411.07522} H. Sahin, H. Akg\"{u}n, Z. B. Siu, S. M. Rafi-Ul-Islam, J. F. Kong, M. B. A. Jalil, and C. H. Lee, 
Protected Chaos in a Topological Lattice, Advanced Science 12, e03216 (2025).

\bibitem{ncommun16-422} K. Sone, M. Ezawa, Z. Gong, T. Sawada, N. Yoshioka, and T. Sagawa, Transition from the topological to the chaotic in the nonlinear Su-Schrieffer-Heeger model, Nat. Commun. 16, 422 (2025).

\bibitem{nphys20-1164} K. Sone, M. Ezawa, Y. Ashida, N. Yoshioka, and T. Sagawa, Nonlinearity-induced topological 
phase transition characterized by the nonlinear Chern number, Nat. Phys. 20, 1164 (2024).

\bibitem{nature596-63} M. J\"{u}rgensen, S. Mukherjee, and M. C. Rechtsman, Quantized nonlinear Thouless pumping, Nature 596, 63 (2021).

\bibitem{PRL128-154101} Q. Fu, P. Wang, Y. V. Kartashov, V. V. Konotop, and F. Ye, Nonlinear Thouless 
Pumping: Solitons and Transport Breakdown, Phys. Rev. Lett. 128, 154101 (2022).

\bibitem{PRL128-113901} M. J\"{u}rgensen and M. C. Rechtsman, Chern Number Governs Soliton Motion in Nonlinear Thouless Pumps, Phys. Rev. Lett. 128, 113901 (2022).

\bibitem{ncommun13-5997} N. Mostaan, F. Grusdt, and N. Goldman, Quantized topological pumping of solitons 
in nonlinear photonics and ultracold atomic mixtures, Nat. Commun. 13, 5997 (2022).

\bibitem{nphys19-420} M. J\"{u}rgensen, S. Mukherjee, C. Jörg, and M. C. Rechtsman, Quantized fractional 
Thouless pumping of solitons, Nat. Phys. 19, 420 (2023).

\bibitem{science384-317} Z. Zhu, M. G\"{a}chter, A.-S. Walter, K. Viebahn, and T. Esslinger, Reversal of quantized Hall 
drifts at noninteracting and interacting topological boundaries, Science 384, 317 (2024).

\bibitem{OL51-1649} R. Li, L. Xu, M. Imran, W. Wang, Y. Jia, and Y. Liu, Symmetry-breaking bifurcation of coupled topological edge states, 
Opt. Lett. 51, 1649 (2026).

\bibitem{PRL42-1698} W. P. Su, J. R. Schrieffer, and A. J. Heeger, Solitons in Polyacetylene, Phys. Rev. Lett. 42, 1698 (1979).

\bibitem{book1} J. K. Asb\'{o}th, L. Oroszl\'{a}ny, and A. P\'{a}lyi, A Short Course on Topological Insulators, 
Vol. 919 (Springer International Publishing, Cham, 2016).

\bibitem{ncommun13-3379} D. Zhou, D. Z. Rocklin, M. Leamy, and Y. Yao, Topological invariant and anomalous edge modes of strongly nonlinear systems, Nat. Commun. 13, 3379 (2022).

\bibitem{FP18-33311} J. Tang, F. Ma, F. Li, H. Guo, and D. Zhou, Strongly Nonlinear Topological Phases of Cascaded Topoelectrical Circuits, Front. Phys. 18, 33311 (2023).

\bibitem{arxiv} K. Sone and Y. Hatsugai, Topological-to-topological transition induced by on-site nonlinearity in a one-dimensional topological insulator, Phys. Rev. Research 8, L012045 (2026).

\bibitem{IEEEJQE29-250} Y. S. Kivshar, Dark solitons in nonlinear optics, IEEE J. Quantum Electron. 29, 250 (1993).

\bibitem{PR298-81} Y. Kivshar, Dark optical solitons: physics and applications, Physics Reports 298, 81 (1998).

\bibitem{PRA103-023503} R. Li and Y. Hadad, Reduced sensitivity to disorder in a coupled-resonator waveguide with disordered coupling coefficients, Phys. Rev. A 103, 023503 (2021).

\bibitem{iscience} W. Wang, Y. Ma, M. Imran, B. Lv, X. He, Y. Jia, H. Tao, Y. Liu, and R. Li, Observation of type-II and type-III Dirac points in circuit lattices, iScience 28, 113778 (2025).

\bibitem{PRB84-195452} P. Delplace, D. Ullmo, and G. Montambaux, Zak phase and the existence of edge states in graphene, Phys. Rev. B 84, 195452 (2011).

\bibitem{AC_limit_1} R. S. MacKay and S. Aubry, Proof of existence of breathers for time-reversible or Hamiltonian networks of weakly coupled oscillators, Nonlinearity 7, 1623 (1994).

\bibitem{AC_limit_2} G. L. Alfimov, V. A. Brazhnyi, and V. V. Konotop, On classification of intrinsic localized modes for the discrete nonlinear Schr\"{o}dinger equation, Physica D: Nonlinear Phenomena 194, 127 (2004).

\bibitem{PD} J. C. Eilbeck, P. S. Lomdahl, and A. C. Scott, The discrete self-trapping equation, Physica D: Nonlinear Phenomena 16, 318 (1985).

\bibitem{bubble_soliton} I. V. Barashenkov and V. G. Makhankov, Soliton-like ``bubbles'' in a system of interacting bosons, Physics Letters A 128, 52 (1988).

\bibitem{PD34-240} I. V. Barashenkov, A. D. Gocheva, V. G. Makhankov, and I. V. Puzynin, Stability of the soliton-like ``bubbles'', Physica D: Nonlinear Phenomena 34, 240 (1989).

\bibitem{PRB103-144202} S. Longhi, Spectral deformations in non-Hermitian lattices with disorder and skin effect: A solvable model, Phys. Rev. B 103, 144202 (2021).

\bibitem{RMP83-247} Y. V. Kartashov, B. A. Malomed, and L. Torner, Solitons in nonlinear lattices, Rev. Mod. Phys. 83, 247 (2011).

\bibitem{CSF157-111912} C. Valagiannopoulos and V. Kovanis, Nonlinear resonances in fast electronic circuits mimicking photonic oscillators, Chaos, Solitons \& Fractals 157, 111912 (2022).

\bibitem{FP20-44204} R. Li, W. Wang, X. Kong, B. Lv, Y. Jia, H. Tao, P. Li, and Y. Liu, Realization of a non-Hermitian Haldane model in circuits, Front. Phys. 20, 44204 (2025).

\bibitem{PRB108-184301} C. Li and Y. V. Kartashov, Topological gap solitons in Rabi Su-Schrieffer-Heeger lattices, Phys. Rev. B 108, 184301 (2023).

\bibitem{LSA14-296} H. Du, H. Zhao, Y. Li, Y. Wang, R. Li, J. Wu, W. Liu, Y. Zhang, L. Xiao, S. Jia, and J. Ma, Observation of nonlinear edge states in 
an interacting atomic trimer array, Light Sci. Appl. 14, 296 (2025).


\end{thebibliography}
\end{document}